\documentclass[sigconf]{acmart}
\usepackage[linesnumbered,ruled]{algorithm2e}
\usepackage{graphicx, subfigure}
\usepackage{url}
\AtBeginDocument{%
  }

\setcopyright{acmlicensed}
\copyrightyear{2024}
\acmYear{2024}
\acmDOI{}

\acmConference[MobiHoc '24]{25th International Symposium on Theory, Algorithmic Foundations, and Protocol Design for Mobile Networks and Mobile Computing}{October 07--10, 2024}{Athens, Greece}

\renewcommand\footnotetextcopyrightpermission[1]{}

\begin{document}

\title{Distributed Scheduling for Throughput Maximization under Deadline Constraint in Wireless Mesh Networks
}

\author{Xin Wang}
\affiliation{%
  \institution{Shanghai Jiao Tong University}
  \city{Shanghai}
  \country{China}}
\email{xin.wang.ece@sjtu.edu.cn}

\author{Xudong Wang}
\affiliation{%
  \institution{Shanghai Jiao Tong University}
  \city{Shanghai}
  \country{China}}
\email{wxudong@ieee.org}


\begin{abstract}
  This paper studies the distributed scheduling of traffic flows with arbitrary deadlines that arrive at their source nodes and are transmitted to different destination nodes via multiple intermediate nodes in a wireless mesh network.
  When a flow is successfully delivered to its destination, a reward will be obtained, which is the embodiment of network performance and can be expressed by metrics such as throughput or network utility.
  The objective is to maximize the aggregate reward of all the deadline-constrained flows, which can be transformed into the constrained Markov decision process (CMDP).
  According to the transformation, a policy gradient-based distributed scheduling (PGDS) method is first proposed, where a primary reward and an auxiliary reward are designed to incentivize each node to independently schedule network resources such as power and subcarriers.
  The primary reward is generated when flows are successfully delivered to their destinations.
  The auxiliary reward, designed based on potential-based reward shaping (PBRS) using local information of data transmission, aims to accelerate the convergence speed.
  Inside this method, a reward feedback scheme is designed to let each node obtain the primary reward.
  Noting that each node selecting resources independently may cause interference and collision which leads to instability of data transmission, a policy gradient-based resource determination algorithm is proposed.
  Moreover, the optimality and convergence of the PGDS method are derived. 
  Especially, when a policy obtained by the algorithm is not matched with the optimal policy but can better deal with the interference, an asymptotic optimum still exists and is further derived.
  The theoretical results are further validated through simulations. 
  Taking throughput as a reward, simulations also show that the PGDS method improves the throughput by at least 70\% and reduces the end-to-end delay.
\end{abstract}

\ccsdesc[500]{Networks~Network algorithms}
\ccsdesc[500]{Theory of computation~Scheduling algorithms}

\keywords{Distributed scheduling, wireless mesh networks, deadline, reward}



\maketitle

\section{Introduction}
Scheduling real-time traffic flows in wireless mesh networks has attracted wide attention due to emerging applications \cite{MoboHoc22Scheduling,9678001,9870687,SIGMETRICS01,3626781}, e.g. in Internet of Things (IoT), autonomous driving, and cyber-physical systems, where time-sensitive packets need to be delivered from their sources to destinations across the whole networks \cite{10234442,9792409,MobiHoc18Multihop}.
When these packets are successfully transmitted to their destinations within the deadline, rewards will be generated. 
The reward is an embodiment of the network performance and can be expressed in terms of throughput or network utility.
The deadline means each flow should be delivered to its destination node on time before a given time requirement, otherwise it will be out-of-date and discarded.
Unlike single-hop networks, where a central node can coordinate the traffic flows and perform scheduling methods~\cite{10125013,9678001,8476232,8486307,MobiHocSignleHop,5462070,6180023}, 
wireless mesh networks are difficult to centrally manage due to the large distance between source-destination pairs.
At the same time, the time-frequency resources in the network are limited, which makes the simultaneous transmission of data flows by non-adjacent links interfere with each other and lead to transmission failure.
Furthermore, different deadline-constrained flows fiercely compete for resources to achieve on-time delivery and thus be coupled with each other, which greatly increases the difficulty of distributed scheduling.
Nowadays, some works of literature have been trying to solve these problems but with little success~\cite{9377564,5340575,2177101,8485769}, such as high-complexity scheduling methods with large overhead and cannot achieve the global optimum.

This paper makes important progress on distributedly scheduling deadline-constrained traffic flows in wireless mesh networks. 
The objective is to maximize the aggregate reward of flows that reach their destinations within their deadlines.
To this end, this paper first formulates the reward maximization problem from a single-packet scheduling perspective considering interference in the network and establishes a relationship between scheduling network resources of packets at each node and the constrained Markov decision process (CMDP).
A policy gradient-based distributed scheduling (PGDS) method based on the relationship is then designed, which quickly adapts to changes in network resources and converges to the global optimum. 
In this method, a primary reward is generated after a packet is successfully delivered to the destination node before its deadline, which is fed back to all intermediate nodes on the end-to-end path and used to guide the scheduling decision. 
Considering the sparsity of the reward generated by the destination node, the lag of feedback information will increase the convergence time.
To improve the convergence speed, an auxiliary reward is designed using potential-based reward shaping (PBRS).
To manage the interference, a resource determination algorithm is proposed, where each node maintains and updates the interference field and avoids collision resources to improve the stability of the PGDS method.
To fully understand the performance of the PGDS method in a theoretical way and clearly know the influence of various parameters in the network, the optimality and convergence properties are derived theoretically as $\mathcal{O}(1/T)$ which is verified to be fast enough to cope with channel variations by simulations.
Finally, taking throughput as a reward, simulation results verify the performance of PGDS under throughput, delay, multiple network topologies, and different channel conditions, and then the maximum throughput are improved by at least 70\% compared with the existing methods.
The main contributions of this paper are summarized as follows.
\begin{itemize}
	\item For wireless mesh networks, the reward maximization model is theoretically established, which is then decomposed to propose a PGDS method.
    The relationship between the PGDS method and CMDP is analyzed in detail, and the method can be effectively used to schedule flows in wireless networks in a distributed manner.
	\item A PGDS method with PBRS is proposed. 
	In this method, a reward feedback scheme is proposed to guide the scheduling on each node, an auxiliary reward based on PBRS is designed to accelerate the convergence, and the resource determination algorithm is proposed to manage the interference.
	\item The optimality and convergence are derived for the proposed PGDS method.
    These theoretical performances are verified and compared with existing methods by simulations.
\end{itemize}

The paper is structured as follows.
In Section~\ref{sec:RelatedWork}, the related work is discussed.
In Section~\ref{sec:NetworkandFramework}, the network model and problem formulation are performed.
In Section \ref{sec:ProblemTransformation}, problem decomposition and the relationship with CMDP are analyzed.
In Section \ref{sec:AlgorithmDesign}, a PGDS method with PBRS is proposed.
In Section \ref{sec:PerformanceAnalysis}, the performance of optimality and convergence are derived.
In Section \ref{sec:SimulationResults}, some important simulation results are shown to illustrate the performance achieved by the proposed method, and the paper is concluded in Section \ref{sec:Conclusion}.


\section{Related Work}\label{sec:RelatedWork}
There is a rich literature on the scheduling of traffic flows with deadline constraints in wireless networks to obtain maximal aggregate rewards, which contains scheduling methods in wireless single-hop networks, wired multi-hop networks, and wireless mesh or multi-hop networks.
The reward can be network utility functions or throughput.

In wireless single-hop networks, where each node can communicate directly with a central node, a common approach is to use the node to coordinate the information of all flows and optimally schedule them under their deadline constraints to obtain maximal utility~\cite{8486307,10125013,9678001,5462070,6180023}. 
An error rate-based prediction model to improve the deadline-constrained throughput is proposed in~\cite{8486307}, a bidding game algorithm between a central node and its neighbors to obtain maximal utility is designed in~\cite{10125013,9678001}, and some centralized methods based on Lyapunov optimization~\cite{5462070,6180023}.
However, these methods are difficult to extend to wireless mesh networks, since it is hard to find a central node that can collect all the information of flows from the whole network.

In wired multi-hop networks, flows are generated at source nodes and transmitted to their destination nodes via multiple intermediate nodes.
If the flow exceeds its deadline, it is invalidated and discarded.
The scheduling methods mainly contain heuristic algorithms without theoretical guarantees~\cite{8726331,6155625}, dynamic programming~\cite{9144447}, a polynomial-time approximation algorithm based on the relaxation of the network utility maximization (NUM) problem~\cite{LIU2019102007}, and distributed scheduling based on CMDP~\cite{8485769}.
Since these methods do not consider interference, they can not be directly extended to wireless mesh networks.
However, a recent work in \cite{8485769}, which developed a connection with CMDP to design distributed scheduling to achieve throughput optimization, inspires us to devise PGDS method. 
It proposed a distributed link selection method for wired multi-hop networks with deadline-constrained flows, where intermediate nodes use a received reward from destination nodes to decide the next-hop transmission link.

In wireless mesh networks, the problem becomes more complicated considering interference.
If the wireless channel is assumed to be stable, some centralized scheduling methods that can achieve maximal performance with deadline constraints can be used~\cite{9705511,9500058,7112191,6155625,10159148}.
Note that these methods require large communication overhead to gather necessary information from all the nodes for a central node to obtain scheduling results.
Distributed scheduling methods are preferred since they allow each node to make local scheduling decisions which is simpler and more efficient.
These methods are mainly divided into three categories.
First, the deadline is used to adjust the scheduling algorithm but not used as a constraint so that the delay can not be guaranteed, such as carrier sense
multiple access (CSMA) scheduling algorithms with delay-based backoff value~\cite{5340575,2177101,6847949,7878682}, back pressure-type controllers adjusted by deadline~\cite{6166337}, opportunistic scheduling with delay differentiation~\cite{5753559}.
Second, the deadline is relaxed to the average delay constraint, such as a
duality-based algorithm~\cite{6619407} and a maximum weight scheduling~\cite{6389738}.
Thirdly, the deadline is constrained but scheduling cannot reach the maximal performance, such as opportunistic scheduling~\cite{5934971} and adaptive CSMA~\cite{9377564}.
Therefore, devising a novel distributed scheduling method is an open and challenging problem.

\section{Network Model and Problem Formulation}\label{sec:NetworkandFramework}
\subsection{Network Model}
A general wireless mesh network is considered in this paper, where $N$ nodes can generate, transmit, and receive flows.
The topology of the network is denoted by a unidirectional graph $G=\{\mathcal{V}, \mathcal{L}\}$, where $\mathcal{V}$ is the set of nodes in the network and $\mathcal{L}$ is the set of links between adjacent nodes.
Neighboring nodes communicate with each other through wireless links. 
There is no central node in the network that can collect information about all nodes and centrally schedule resources.
Considering that orthogonal frequency division multiplexing (OFDM) is widely used in 6G and beyond Wireless networks, the subcarrier and time slot that can be scheduled are defined as the network resources in this paper.
Therefore, nodes at each time slot can perform the scheduling of subcarriers to transmit flows.
Each node is in half-duplex mode, that is, it can only receive or transmit flows in the same time slot.
$F$ real-time flows are generated at their source nodes and transmitted to their destination nodes through multiple intermediate nodes with each flow having end-to-end deadline constraint $\tau_f$.
The weight of flow $f$ is $w_f$, which is determined by its priority.
While transmitting flows in any time slot, each node needs to select subcarriers and transmission power for flows at that node.
The total subcarriers in the network is $C$ and the maximal transmission power at each node is $P$.
The channel gain from node $i$ to node $j$ is $h_{ij}$ ($0\leq h_{ij} \leq 1$). 
Wireless interference can prevent non-adjacent links from transmitting flows at the same time, which increases the difficulty of scheduling.
For any node, it interferes with those nodes that can receive its signal power.
Interference field of node $i$ is $\mathcal{I}_i(t)$, which is defined as $\mathcal{I}_i(t)=\{j|j\in \mathcal{V},P_ih_{ij}^2>P_{\rm Th}\}$.
$P_i$ is the transmission power at node $i$ and $P_{\rm Th}$ is the detection threshold of interference.
To avoid transmission failure due to interference, all links that are affected by node $i$ cannot use more than $C$ subcarriers.

\subsection{Reward Maximization Problem}
\begin{figure}[h]
  \centering
  \includegraphics[width=\linewidth]{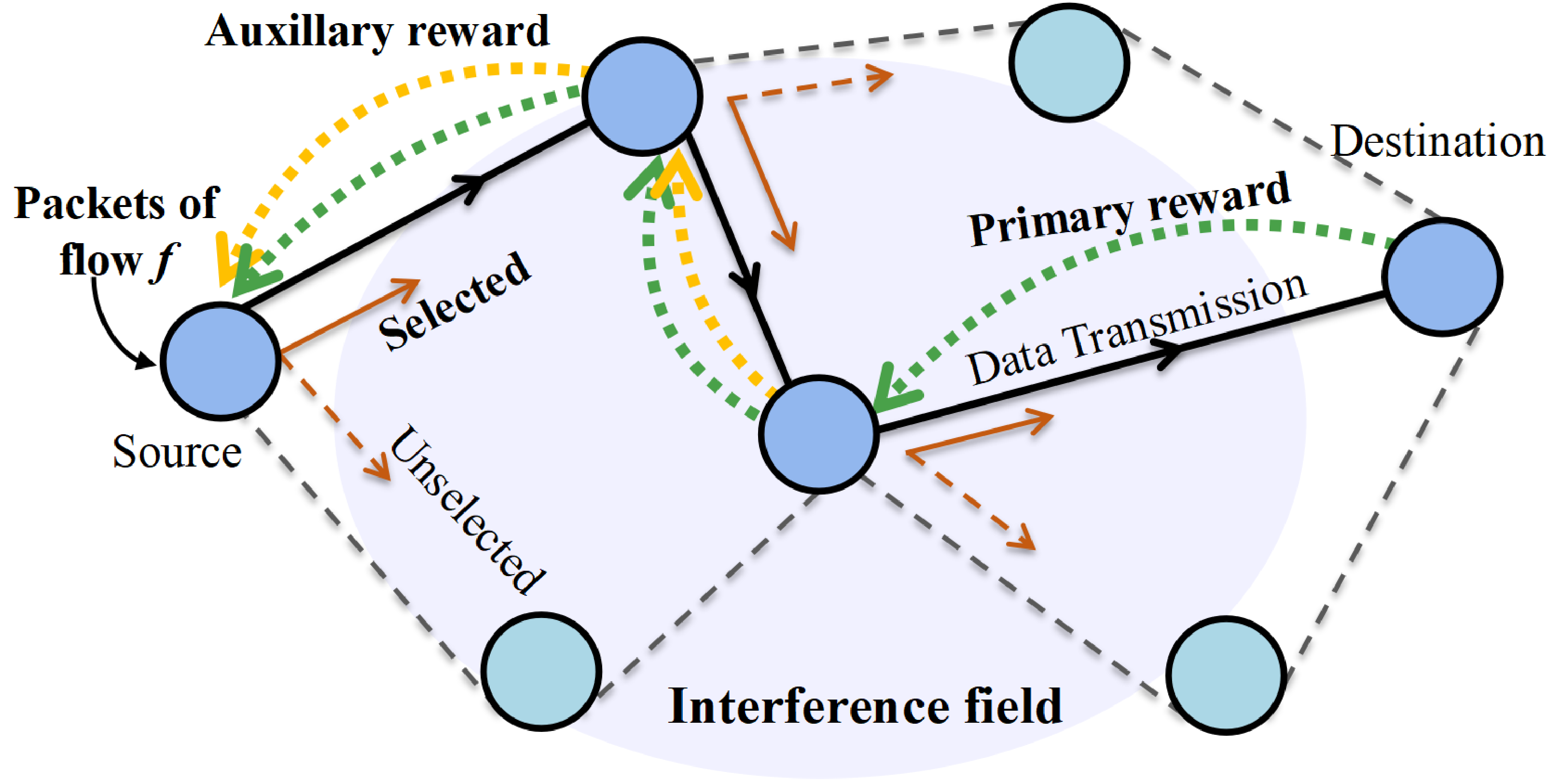}
  \caption{Distributed scheduling for packets of flow $f$.}
  \label{fig:scheme}
\end{figure}

Note that any flow $f$ is composed of packets, and each packet has a deadline $\tau_f$.
Distributed scheduling for any packet of flow $f$ is shown in Fig.~\ref{fig:scheme}.
A primary reward $u(t)$ is generated after the packet is successfully delivered to the destination node within the deadline.
$u(t)$ is a function of the delivered packet, such as throughput or utility functions.
Define a unit packet to be denoted by $\sigma$ and the set of all unit packets of flow $f$ to be denoted by $\mathcal{P}_f(t)$ ($\sigma\in\mathcal{P}_f(t)$).
Without loss of generality, all flows in a network can be represented as a combination of multiple unit packets.
The wireless mesh network considered in this paper is a time-slotted system, the number of subcarriers and transmission power need to be selected for each unit packet at each time slot.
When a unit packet is transmitted from the source node to the destination node, its position (the node $i$ where it is located) and time-till-deadline (TTD) $\tau$ are constantly changing. 
These two varying parameters are used to define the state $s_{\sigma}(t)$ of the packet. 
$s_{\sigma}(t)\in \mathcal{S}$ and $\mathcal{S}$ is denoted as
\begin{align}
    \mathcal{S} = \left\{(i,\tau)|i\in\mathcal{V},0\le\tau\le\tau_f\right\}.
\end{align}
Scheduling the packet includes determining its next transmission node $j$ and the number $n_{\rm c}$ and power $n_{\rm p}$ of subcarriers, which is denoted as $a_{\sigma}(t)$.
$a_{\sigma}(t)\in \mathcal{A}$ and $\mathcal{A}$ is given by
\begin{align}
    \mathcal{A} = \left\{(j,n_{\rm c},n_{\rm p})|j\in\mathcal{V}_i,0\le n_{\rm c}\le C,0\le n_{\rm p} \le P\right\},
    \label{eq:action}
\end{align}
where $\mathcal{V}_i$ denotes the set of neighbor nodes of node $i$, which is also a subset of $\mathcal{V}$.
A scheduling policy $\pi[a_{\sigma}(t)|s_{\sigma}(t)]$ is defined as the probability of choosing schedule $a_{\sigma}(t)$ in state $s_{\sigma}(t)$.
The policy is equal to 1 means definitely selects $a_{\sigma}(t)$.
When the packet is successfully delivered to the destination node, the reward is generated, which is expressed as
\begin{align}
    r_{\rm D}\left[s_{\sigma}(t),a_{\sigma}(t),s_{\sigma}(t+1)\right]=\left\{\begin{matrix}
                u(t),  j \text{ is destination and } \tau\ge0\\
                0,  \text{otherwise}.
                \end{matrix}\right.
\end{align}
The objective is to maximize the aggregate rewards of all the packets in the network.
However, $r_{\rm D}(\cdot)$ can only be obtained at the destination, which does not reflect the impact of the scheduling of packets in intermediate nodes and makes it difficult to determine the scheduling of intermediate nodes.
Therefore, an auxiliary reward $\mathcal{H}\left[s_{\sigma}(t),a_{\sigma}(t),s_{\sigma}(t+1)\right]$ needs to be designed to reflect the global impact of the scheduling of intermediate nodes, and the new reward can be expressed as
\begin{align}\notag
    &r\left[s_{\sigma}(t),a_{\sigma}(t),s_{\sigma}(t+1)\right]\\
    &=r_{\rm D}\left[s_{\sigma}(t),a_{\sigma}(t),s_{\sigma}(t+1)\right]+\mathcal{H}\left[s_{\sigma}(t),a_{\sigma}(t),s_{\sigma}(t+1)\right].
    \label{eq:newReward}
\end{align}
Using the new reward, the expected cumulative rewards of the packet with state $s_{\sigma}(t)$ is
\begin{align}
    R_{r}^{\pi}\left[s_{\sigma}(t)\right] = \mathbb{E}_{\pi}\left\{\sum_{k=0}^{\tau_f-1}r\left[s_{\sigma}(t+k),a_{\sigma}(t+k),s_{\sigma}(t+k+1)\right]\right\}.
    \label{eq:ECReward}
\end{align}
Distinguished from the auxiliary reward, $r_{\rm D}(\cdot)$ is called the primary reward.
It should be emphasized that the role of the auxiliary reward is to facilitate the scheduling of packets by intermediate nodes, and it must not influence the scheduling policy $\pi$ that maximizes the primary reward.
This means that the scheduling policy obtained using the new reward is the same as the optimal scheduling policy obtained with the primary reward, that is, the policy-invariant auxiliary reward needs to be designed.

Moreover, constraints on the number and power of subcarriers need to be established.
Define $c\left[s_{\sigma}(t),a_{\sigma}(t),s_{\sigma}(t+1)\right]$ is the number of subcarriers and $p[s_{\sigma}(t),$ $a_{\sigma}(t),s_{\sigma}(t+1)]$ is the power selected from state $s_{\sigma}(t)$ to $s_{\sigma}(t+1)$, which are equal to $n_{\rm c}$ and $n_{\rm p}$ in $a_{\sigma}(t)$ in \eqref{eq:action}, respectively.
The expected number of subcarriers under the policy $\pi$ is 
\begin{align}
    D_{\rm c}^{\pi}\left[s_{\sigma}(t)\right]=\mathbb{E}_{\pi}\left\{c\left[s_{\sigma}(t),a_{\sigma}(t),s_{\sigma}(t+1)\right]\right\}.
    \label{eq:subcarrierCons}
\end{align}
The expected power under the policy $\pi$ is 
\begin{align}
    D_{\rm p}^{\pi}\left[s_{\sigma}(t)\right]=\mathbb{E}_{\pi}\left\{p\left[s_{\sigma}(t),a_{\sigma}(t),s_{\sigma}(t+1)\right]\right\}.
    \label{eq:powerCons}
\end{align}
Considering the constraint on the number of subcarriers and the transmit power of each node, the problem of maximizing the aggregate new reward of all the packets is formulated as
\begin{align}
\label{eq:MIOP}
\max_{\pi}&\lim_{T\to \infty} \frac{1}{T} \sum_{t=0}^{T-1}\sum_{f\in\mathcal{F}}\sum_{\sigma\in \mathcal{P}_f(t)}w_fR_{r}^{\pi}\left[s_{\sigma}(t)\right],\\
&\lim_{T\to \infty} \frac{1}{T} \sum_{t=0}^{T-1}\sum_{f\in\mathcal{F}}\sum_{\sigma\in \mathcal{P}_{fi}(t)\cup \mathcal{P}_{\mathcal{I}_i}(t)}D_{\rm c}^{\pi}\left[s_{\sigma}(t)\right]\leq C, \forall i\in \mathcal{V},\\
&\lim_{T\to \infty} \frac{1}{T} \sum_{t=0}^{T-1}\sum_{f\in\mathcal{F}}\sum_{\sigma\in \mathcal{P}_{fi}(t)}D_{\rm p}^{\pi}\left[s_{\sigma}(t)\right]\leq P, \forall i\in \mathcal{V},
\end{align}
where $\mathcal{P}_{fi}$ represents the set of packets of flow $f$ at node $i$ and $\mathcal{P}_{\mathcal{I}_i}$ represents the set of packets in the interference field of node $i$.

\section{Problem Decomposition}\label{sec:ProblemTransformation}
Note that the subcarrier in \eqref{eq:MIOP} is a discrete variable and the power is a continuous variable, the objective function can be either discrete or continuous, thus this is a mixed integer optimization problem (MIOP).
To solve the MIOP in a distributed way, the relationship between different states needs to be clarified first.
Take the transmission of a unit packet of flow $f$ as an example.
In wireless mesh networks, a unit packet at state $s_{\sigma}(t)=(i,\tau)$ selects a scheduling way $a_{\sigma}(t)=(j,n_{\rm c},n_{\rm p})$ and wants to transmit to another state $s_{\sigma}(t+1)=(j,\tau-1)$.
This packet is not necessarily able to transmit to node $j$, but has a transition probability ${\rm Pr}(s_{\sigma}(t+1)|s_{\sigma}(t),a_{\sigma}(t))$.
This probability is related to the channel characteristics of the current transmission link and the transmission technology of the physical layer, such as the coding and modulation methods.
In the paper, the probability is expressed as ${\rm Pr}(SINR_{ij}\geq SINR_{\rm Th}){\rm Pr}_{\rm e}(SINR_{ij})$, where $SINR_{ij}$ means the signal-to-interference-noise ratio (SINR) at node $j$ and is denoted as
\begin{align}
    SINR_{ij}=\frac{p\left[s_{\sigma}(t),a_{\sigma}(t),s_{\sigma}(t+1)\right]h_{ij}^2}{N_0p\left[s_{\sigma}(t),a_{\sigma}(t),s_{\sigma}(t+1)\right]\Delta_c+P_{\mathcal{I}_i}},
\end{align}
where $N_0$ is the power spectral density of additional white Gaussian noise (AWGN), $\Delta_c$ is the subcarrier space, and $P_{\mathcal{I}_i}$ is the interference power from interference field $\mathcal{I}_i$.
$P_{\mathcal{I}_i}=\sum_{\sigma\in\mathcal{P}_{\mathcal{I}_i}}p[s_{\sigma}(t),a_{\sigma}(t),$ $s_{\sigma}(t+1)]h_{kj}^2$ where $k$ is the node in the interference field $\mathcal{I}_i$.
${\rm P}_{\rm e}$ means the packet error rate under specific coding and modulation methods.
Note that this transition probability is only related to the current link and has nothing to do with other links that the packet passes through, that is
\begin{align}\notag
    {\rm Pr}\left[s_{\sigma}(t+1)|s_{\sigma}(0),a_{\sigma}(0),s_{\sigma}(1),a_{\sigma}(1),...,s_{\sigma}(t),a_{\sigma}(t)\right] \\
    = {\rm Pr}\left[s_{\sigma}(t+1)|s_{\sigma}(t),a_{\sigma}(t)\right].
\end{align}
Therefore, the process of transmitting the packet from state $s_{\sigma}(t)$ to the next state $s_{\sigma}(t+1)$ is independent of the historical state, that is, it conforms to the Markov property.

It is stated above that the MIOP in \eqref{eq:MIOP} can be viewed as the constrained Markov decision process (CMDP), and then the Lagrangian decomposition can be used.
The detailed derivation process is in Appendix A in \cite{DDLArt} and the main result after the decomposition is
\begin{align}
    \mathfrak{L}\left ( \pi,\lambda,\mu \right )=\lim_{T\to \infty} \frac{1}{T} \sum_{t=0}^{T-1}\sum_{f\in\mathcal{F}}\sum_{i\in\mathcal{V}}\sum_{\sigma\in \mathcal{P}_{fi}(t)}R_{r_{\rm L}}^{\pi,\lambda,\mu}\left[s_{\sigma}(t)\right],
    \label{eq:Lagrangian}
\end{align}
where the constant term is ignored, $\lambda$ and $\mu$ are the Lagrange multipliers, and $R_{r_{\rm L}}^{\pi,\lambda,\mu}\left[s_{\sigma}(t)\right]$ is to replace $r(\cdot)$ in \eqref{eq:ECReward} with $r_{\rm L}(\cdot)$.
$r_{\rm L}(\cdot)$ is the reward under the Lagrange multipliers and is derived as
\begin{align}
    r_{\rm L}^{\sigma}(t) = w_fr^{\sigma}(t)-\lambda_ic^{\sigma}(t)-\mu_{i}p^{\sigma}(t)-\frac{\lambda_i}{N_i}\sum_{\sigma^{\prime}\in\mathcal{P}_{\mathcal{I}_i}(t)}c^{\sigma^{\prime}}(t),
    \label{eq:RewardL}
\end{align}
where $r^{\sigma}(t)$, $c^{\sigma}(t)$, and $p^{\sigma}(t)$ are simplified representations of $r(\cdot)$ in \eqref{eq:ECReward}, $c(\cdot)$ in \eqref{eq:subcarrierCons}, and $p(\cdot)$ in \eqref{eq:powerCons}.
$N_i$ is the number of packets at node $i$.
It's worth noting that $r_{\rm L}^{\sigma}(t)$ can be viewed as two parts.
One part is the reward obtained for successful transmission, which is the first three terms on the right-hand side of \eqref{eq:RewardL}.
The other part is the effect brought by the transmission of other packets in the interference field, which is the fourth term on the right side of \eqref{eq:RewardL}.
Recall the Lagrangian in \eqref{eq:Lagrangian}, it decomposes the packet scheduling to each node, that is, the packets of each node can obtain and constrain its own reward through the interference field, and then these rewards can be used to achieve distributed scheduling.

For the above-transformed problems, CMDP provides some standard algorithmic frameworks for solving them, such as value-based and policy-based algorithms~\cite{AltemanCMDP,ValueBasedICML,NEURIPS2020_5f7695de}.
However, these algorithms cannot be directly used to solve the transformed problem, and the two main challenges are as follows.
One is that the feasible region is changing. 
Since the interference range changes with scheduling and channel gain, the number of available subcarriers is constantly changing.
Therefore,  when the scheduling policy is updated, the feasible region also needs to be updated synchronously.
The other is the choice of scheduling policy. 
The collision occurs when the total number of subcarriers selected by packets in the interference field exceeds $C$. 
In the actual scheduling, it is necessary to avoid this collision situation, so as to reduce the probability of transmission errors and improve the stability of the network.
Note that the distributed scheduling framework using value-based algorithms proposed in wired multi-hop networks inspires the design in this paper~\cite{8485769}.
However, this framework cannot be applied to wireless multi-hop networks since it does not take into account the effects of wireless interference.
Considering that the value-based algorithm may converge to a local optimum and it is difficult to find a global optimal policy when the action space and state space are large, policy-based algorithm framework is adopted in this paper.
Therefore, this paper proposes a policy gradient-based distributed scheduling (PGDS) method and details are designed in section \ref{sec:AlgorithmDesign}.

\section{Scheme and Algorithms in PGDS method with PBRS}\label{sec:AlgorithmDesign}
In the PGDS method, when a packet reaches a state, it selects the next-hop link, number of subcarriers, and power of the next-hop transmission according to Algorithm~\ref{alg:algorithm1}.
The packet then listens for the selection of other states in the interference field and determines whether the transmission needs to be performed.
When the packet is successfully delivered to the destination node, the destination node will generate a reward.
The reward is then fed back to each node along the end-to-end path and used to update scheduling policies according to a reward feedback scheme in section~\ref{sec:RewardFeedbackScheme}. 
When future packets arrive at these nodes, the number of subcarriers and power can then be selected based on the scheduling policy.
The auxiliary reward, which is based on PBRS and can guarantee the policy optimality is derived in section~\ref{sec:AuxDesign}.
In summary, based on this method, any packet at any node in the network can decide the resource to transmit in the next hop according to the scheduling policy, so as to realize distributed scheduling.
The PGDS framework consists of three key parts: reward feedback scheme, auxiliary reward design, and policy gradient-based resource determination algorithm.
The main design methods for these parts are described in detail below.

\subsection{Reward Feedback Scheme}
The key to performing the PGDS method is that the rewards generated by the destination node need to affect all nodes along the end-to-end path.
This effect can be described by the derivation of $R_{r_{\rm L}}^{\pi,\lambda,\mu}$ in \eqref{eq:Lagrangian}, which is denoted by
\begin{align}\notag
    &R_{r_{\rm L}}^{\pi,\lambda,\mu}[s_{\sigma}(t)]= \sum_{a_{\sigma}(t)\in\mathcal{A}_i}\pi[a_{\sigma}(t)|s_{\sigma}(t)]\left\{{\rm Pr}(\cdot)\right.\\
    &\left.\left[r^{\sigma}_{\rm L}(t)+R_{r_{\rm L}}^{\pi,\lambda,\mu}[s_{\sigma}(t+1)]\right]+[1-{\rm Pr}(\cdot)]R_{r_{\rm L}}^{\pi,\lambda,\mu}[s_{\sigma}^{\prime}(t+1)]\right\},
\end{align}
where ${\rm Pr}(\cdot)={\rm Pr}[s_{\sigma}(t+1)|s_{\sigma}(t),a_{\sigma}(t)]$, $s_{\sigma}(t)=(i,\tau)$, $s_{\sigma}(t+1)=(j,\tau-1)$, and $s_{\sigma}^{\prime}(t)=(i,\tau-1)$.
The above derivation shows that in order to know the global impact of scheduling subcarriers and power in a state, it is necessary to know the expected cumulative reward in the next state, and so on until the end of the destination node.
This means that the destination node needs to feed back the generated reward so that every intermediate node knows the global impact of its schedule.

\begin{figure}[h]
  \centering
  \includegraphics[width=.9\linewidth]{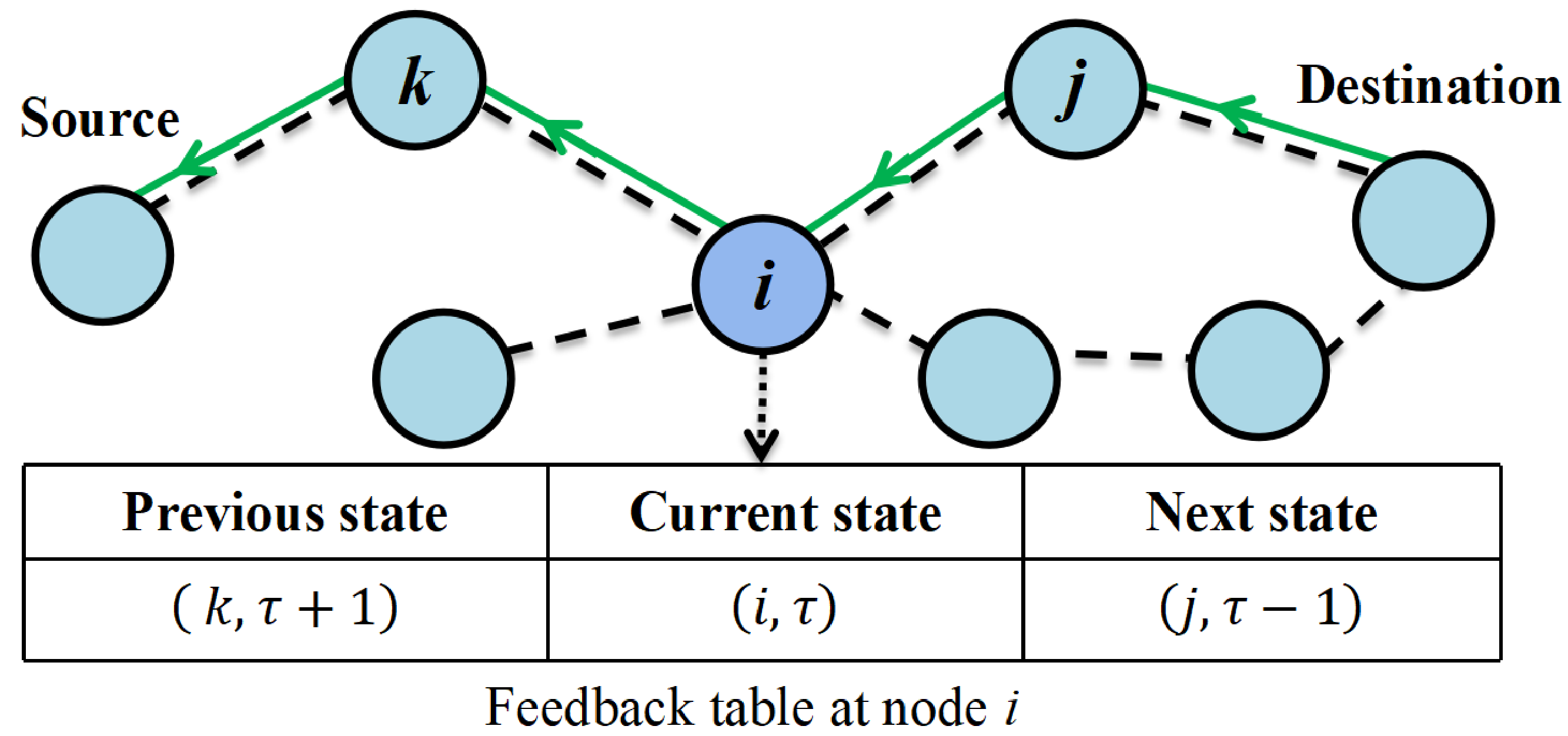}
  \caption{Reward feedback at node $i$.}
  \label{fig:feedback}
\end{figure}

The reward feedback scheme is designed to realize the above process as shown in Fig.~\ref{fig:feedback}.
The scheme includes the establishment of a feedback table and the backtracking of rewards.
When packets are transmitted from the source node to the destination node, feedback tables are created on each node.
The feedback table records the current state, previous state, and next state of the packet shown in Fig.~\ref{fig:feedback}.
Since the transfer between states has Markov property, these three states can effectively help the reward backtracking.
More specifically, the transmission of packets between different nodes corresponds to the process of state transfer, and only packets belonging to the same flow can be transferred to each other.

When the packet is successfully delivered to the destination node, the reward is generated according to \eqref{eq:RewardL}.
The reward is then fed back to each node along the end-to-end path and the basic process is described below. 
The reward needs to carry information about which flow it belongs to and the state of the last hop.
For example, the reward at node $j$ carries information: flow $f$, state $(j,\tau-1)$.
When node $i$ receives a reward from node $j$ and needs to find its next feedback node in Fig.~\ref{fig:feedback}, node $i$ needs to check the feedback table.
According to the information carried by the reward, node $i$ first matches the "Next state" to this information.
Since the time of one-hop transmission is a unit time slot, node $i$ can infer the "Current state" of the reward should be $(i,\tau)$.
After matching the "Next state" and "Current state" in the feedback table, node $i$ can finally find the "Previous state".
The identifier of the node in the previous state is the next feedback node for the reward.
In order not to make the size of the feedback table too large, when the reward determines the node that needs feedback, the information belonging to the feedback path needs to be cleaned out at the node in time.
Therefore, according to the above reward feedback process, the reward generated by the destination node can affect every node on the end-to-end path.
\label{sec:RewardFeedbackScheme}

Note that this section designs the process of reward feedback from the perspective of scheme design. 
In practice, it is necessary to design a specific signaling transmission protocol to ensure the effective operation of the scheme, which will be studied in the future.

\subsection{Auxiliary Reward Design based on PBRS}
\label{sec:AuxDesign}
Since only the destination node can generate a reward, it takes some time for this reward to feedback to the intermediate node. 
During this period, the intermediate nodes lack the information to guide the scheduling, which makes the convergence speed slow.
To speed up the convergence, this paper designs an auxiliary reward to assist the reward generated at the destination to complete fast scheduling.
Note that the auxiliary reward should not affect the optimality of the scheduling policy, and only the information that can be collected locally can be exploited.
Potential-based reward shaping (PBRS) has been proven to improve the convergence without changing the optimality, but how to design an effective reward shaping function needs to combine with the specific network situation.
First, the representation of an auxiliary reward needs to be designed.
According to PBRS, it has the following result~\cite{3041938,RewardShaping}.

\textbf{Lemma 1}
Recall that the new reward in \eqref{eq:newReward}, the policy-invariant auxiliary reward should be formatted as
\begin{align}
    \mathcal{H}[s_{\sigma}(t),a_{\sigma}(t),s_{\sigma}(t+1)] = \phi[s_{\sigma}(t+1)]-\phi[s_{\sigma}(t)],
\end{align}
where $\phi(\cdot)$ is a potential function and belongs to $\mathbb{R}$.

Using the format of auxiliary reward in Lemma 1, the optimal policy will not be changed.
Then, we need to determine the metric that will be used for designing the potential function.
Noting that the primary reward in \eqref{eq:newReward} produced by the destination node is a function of the successfully delivered packet, the auxiliary reward should also have a metric related to it.
The number of packets received by the destination node is related to the number of packets transmitted by the intermediate node. 
The fewer the number of packets transmitted by the intermediate node, the fewer the number of packets successfully delivered to the destination node.
From this perspective, the number of packets transmitted by the intermediate node is the bottleneck that limits the number of packets delivered by the destination node.
Therefore, the link that transmits a large number of packets should be selected as far as possible, that is, the number of packets in the same state should be used as a metric.
This metric is denoted as $u_{\rm L}[s_{\sigma}(t)]$.
In addition, to deliver the packet as soon as possible, it is necessary to select the intermediate node as close to the destination node as possible during scheduling, so the minimum number of hops from the intermediate node to the destination node should be used as a metric.
This metric is denoted as $d_{\sigma i}$ where $i$ is the same as that in $s_{\sigma}(t)$.
To ensure the consistency of convergence, the potential function needs to have a consistent form with the primary reward, and the potential function also needs to have a tendency to change to the global optimum, that is, the maximum potential is at the destination node and the minimum potential is at the source node.
Finally, the potential function is designed as

\textbf{Theorem 1} Suppose the format of an auxiliary reward in Lemma 1, the auxiliary reward in wireless mesh networks is designed as
\begin{align}\notag
    \mathcal{H}[s_{\sigma}(t),a_{\sigma}(t),s_{\sigma}(t+1)]=\frac{u_{\rm L}\left[s_{\sigma}(t+1)\right]}{d_{\sigma j}}-\frac{u_{\rm L}\left[s_{\sigma}(t)\right]}{d_{\sigma i}},
\end{align}
where $j$ is the same as that in $s_{\sigma}(t+1)$.

$d_{\sigma i}$ and $d_{\sigma j}$ can be easily obtained by a simple distributed signaling interaction protocol in~\cite{Gradient}.
The proof of policy invariance with this potential function is derived in Appendix B in \cite{DDLArt}.

\begin{algorithm}[t]
    \caption{Policy Gradient-based Resource Determination.}
    \label{alg:algorithm1}
    \SetKwComment{Comment}{/* }{ */}
    \KwIn{$R_{r_{\rm L}}^{\pi,\lambda,\mu}[s_{\sigma}(t-1)]$, policy $\pi_{\theta_{sa}}(t-1)$, $n_{\sigma}$.}
    \KwOut{$R_{r_{\rm L}}^{\pi,\lambda,\mu}[s_{\sigma}(t)]$, $\pi_{\theta_{sa}}(t)$, $\tilde{c}\left[s_{\sigma}(t),a_{\sigma}(t),s_{\sigma}(t+1)\right]$, and $\tilde{p}\left[s_{\sigma}(t),a_{\sigma}(t),s_{\sigma}(t+1)\right]$.}  
    \BlankLine
 
    \For{$s_{\sigma}(t)\in\mathcal{S}_i$ and $t>0$}{
    Calculate $\left.\nabla_{\theta}\mathfrak{L}\left ( \pi,\lambda,\mu \right )\right|_{s_{\sigma}(t-1),a_{\sigma}(t-1)}$.

    Policy gradient update:
    \begin{align}\notag
        \pi_{\theta_{sa}}(t)=\frac{\pi_{\theta_{sa}}(t-1)e^{\eta_1 \left.\nabla_{\theta}\mathfrak{L}\left ( \pi,\lambda,\mu \right )\right|_{s_{\sigma}(t-1),a_{\sigma}(t-1)}}}{\sum_{a\in \mathcal{A}_i}\pi_{\theta_{sa}}(t)e^{ \eta_1 \left.\nabla_{\theta}\mathfrak{L}\left ( \pi,\lambda,\mu \right )\right|_{s_{\sigma}(t-1),a}}}
    \end{align}

    Scheduling $a_{\sigma}(t)$ with the probability $\pi_{\theta_{sa}}(t)$: obtain $c^{\sigma}(t)$ and $p^{\sigma}(t)$.
    
    Calculate the number of bits that can be transmitted under the channel gain $h_{ij}$: $\tilde{n}=c^{\sigma}(t)\Delta_c\log_2\left(1+\frac{p^{\sigma}(t)h_{ij}^2}{N_0c^{\sigma}(t)\Delta_c}\right)\Delta_t$.
    
    \While{$|\tilde{n}-n_{\sigma}|>n_{\epsilon }$ or $\tilde{n}\ne 0$}
    {
    Repeat the above probabilistic scheduling and calculate $\tilde{n}$.

    \If{the number of repetitions reaches the limit}
    {
    $c^{\sigma}(t)=0$ and $p^{\sigma}(t)=0$.
    }
    }
    }


    
    Calculate total number of subcarriers selected in node $i$ and $\mathcal{I}_i$: \Comment*[r]{Capacity unavailable condition}
    \begin{align}\notag
        C_i = \sum_{f\in\mathcal{F}}\sum_{\sigma\in \mathcal{P}_{fi}(t)  \cup \mathcal{P}_{\mathcal{I}_i}(t)}c^{\sigma}(t)
    \end{align}

    
    \If{$C_i>C$}
    {
    Remove the scheduling with the smallest $R_{r_{\rm L}}^{\pi,\lambda,\mu}[s_{\sigma}(t-1)]$ until $C_i\le C$.

    Final resource for each state is $\tilde{c}^{\sigma}(t)$ and  $\tilde{p}^{\sigma}(t)$.
    }
    
    Using the selected resources to update Lagrangian multipliers:
        \begin{align} \notag
        \lambda_{i}(t+1) = \left(\lambda_{i}(t) + \eta_2 \left\{C-\sum_{f\in\mathcal{F}}\sum_{\substack{\sigma\in \mathcal{P}_{fi}(t) \\ \cup \mathcal{P}_{\mathcal{I}_i}(t)}}D_{\tilde{c}}^{\pi}\left[s_{\sigma}(t)\right]\right\}\right)_{+},
        \end{align}

        \begin{align} \notag
        \mu_{i}(t+1) = \left(\mu_{i}(t) + \eta_3 \left\{P-\sum_{f\in\mathcal{F}}\sum_{\sigma\in \mathcal{P}_{fi}(t)}D_{\tilde{p}}^{\pi}\left[s_{\sigma}(t)\right]\right\} \right)_{+}.
        \end{align}

\end{algorithm}

\subsection{Determination of Resources}\label{sec:IFRSAlg}
When one node receives feedback information using the reward feedback scheme in section~\ref{sec:RewardFeedbackScheme}, it can calculate the scheduling policy $\pi$ and select the number and power of subcarriers.
Note that the scheduling space is composed of the number and power of subcarriers and the next-hop transmission node. 
The power can be divided into different levels and it is a discrete value.
The number of subcarriers may be large in wideband communication scenarios, which means that the size of the scheduling space may be large.
However, most of the policy-based methods proposed to solve a CMDP problem are not yet applicable to such a large space.
Policy gradient-based methods have the ability to deal with such a large scheduling space, and have the theoretical global convergence guarantee.
Therefore, a policy gradient-based framework is used to design the resource determination algorithm in Algorithm \ref{alg:algorithm1}.
In this framework, a parameterized policy $\pi_{\theta} (\theta\in\mathbb{R})$ is defined as $\pi_{\theta}(a|s)=\frac{\exp(\theta_{sa})}{\sum_{a^{\prime}\in \mathcal{A}}\exp(\theta_{sa^{\prime}})}$, which is differentiable and tractable.
The policy gradient of the Lagrangian in \eqref{eq:Lagrangian} is derived in Appendix C in \cite{DDLArt} and the main result is
\begin{align}\notag
    \nabla_{\theta}\mathfrak{L}\left ( \pi,\lambda,\mu \right ) = &\lim_{T \to \infty} \frac{1}{T}\sum_{f\in\mathcal{F}}\sum_{i\in\mathcal{V}}\sum_{\sigma\in \mathcal{P}_{fi}(t)}\nabla_{\theta}\log \pi_{\theta}[a_{\sigma}(t)|s_{\sigma}(t)]\\
    &\left\{D^{\pi,\lambda,\mu}_{r_{\rm L}}[s_{\sigma}(t),a_{\sigma}(t)]-R^{\pi,\lambda,\mu}_{r_{\rm L}}[s_{\sigma}(t)]\right\},
    \label{eq:PolicyGradient}
\end{align}
where $D^{\pi,\lambda,\mu}_{r_{\rm L}}[s_{\sigma}(t),a_{\sigma}(t)]$ is the expected cumulative reward obtained under the given $s_{\sigma}(t)$ and $a_{\sigma}(t)$.
The parameterized gradient is given by 
\begin{align}
    \theta_{sa}(t+1) = \theta_{sa}(t)+\eta_1 \left.\nabla_{\theta}\mathfrak{L}\left ( \pi,\lambda,\mu \right )\right|_{s_{\sigma}(t),a_{\sigma}(t)},
\end{align}
where $\eta_1$ is a step size.
Consider the relationship between $\pi_{\theta}(a|s)$ and $\theta_{sa}$, the policy gradient is obtained and shown in Algorithm \ref{alg:algorithm1}.
The algorithm is used at any node $i$, $n_{\sigma}$ denotes the bit number of the unit packet, $()_{+}$ denotes the projection onto a positive real set $\mathbb{R}_{+}$, $\Delta_t$ denotes the unit time slot, and $s_{\sigma}(t)=(i,\tau)$.

Note that when the number of subcarriers scheduled by all states within the interference range exceeds $C$, the collision of packet transmission will occur, which leads to packet transmission failure.
Since there is no reward for this packet transmission failure, it is actually equivalent to the packet giving up the transmission at the current time slot, so Algorithm~\ref{alg:algorithm1} uses the capacity unavailable condition in line 13 to make these packets give up the transmission in advance.
When these packets finish scheduling resources independently, they will broadcast their selection result and the reward value in the history to all the nodes in the interference field.
If a packet finds that the sum of all subcarriers in its interference range exceeds $C$, it will compare the reward values, and let the scheduling with the minimum reward value give up the transmission in this time slot, so as to ensure no interference.
The above resource determination process needs to use a distributed signaling interaction protocol, which can be quickly implemented by some existing methods \cite{9172113,6265054}, such as nucleus-based methods, which can allocate the subcarriers required by each packet to orthogonal frequencies.
The non-interference transmission will make some packets give up the transmission in advance, the state of these packets is also converted at the current node, and no reward will be obtained. 
Therefore, the same effect as the interference-based transmission will not affect the convergence of the distributed method to the optimal scheduling policy.
However, this will lead to a change in scheduling policy and bring some errors, and detailed analysis and derivation are shown in Appendix E in \cite{DDLArt}.

\section{Optimality and Convergence}\label{sec:PerformanceAnalysis}
This section derives the optimality and convergence properties of the PGDS method.
Considering that the algorithm in section \ref{sec:IFRSAlg} changes the scheduling results under capacity unavailable condition, it may lead to an error from the optimal scheduling.
Therefore, this section analyzes the optimality without the capacity unavailable condition first, and then considers the error caused by changing the scheduling results to further derive the optimality. 

\subsection{Without Capacity Unavailable Condition}
Define the optimal policy as $\pi^{*}$ and the optimal expected cumulative reward for each packet is $R^{*}_{r\sigma}$.
The relationship between the optimality of the PGDS method and the network parameters is crucial, as this facilitates the dynamic management of the network.
The network parameters considered in this paper include the number of nodes $N$, the number of flows $F$, the number of subcarriers $C$, and the maximum deadline constraint $\tau_{\max}$.
The global convergence of PGDS method is derived in Appendix D in \cite{DDLArt}, and the main result is shown below.

\textbf{Theorem 2} Without capacity unavailable condition, each state will directly perform the selected number and power of subcarriers.
The optimality gap between the objective and optimal aggregate reward $R^{*}_{r}$ is
\begin{align}\notag
    \frac{1}{T} \sum_{t=0}^{T-1}\sum_{f\in\mathcal{F}}\sum_{\sigma\in \mathcal{P}_f(t)}  w_f \left\{ R^{\pi}_{r}\left[s_{\sigma}(t)\right ]-R^{*}_{r\sigma} \right\}\leq 
    \frac{g(N,F,C,\tau_{\max})}{T},
\end{align}
where $g$ is the function of the network parameters shown in Appendix D in \cite{DDLArt}, and it also has a limited value.

Since $g$ is limited, if the iteration time $T$ is large enough, it is obtained that
\begin{align}
    \lim_{T \to \infty}\frac{g(N,F,C,\tau_{\max})}{T}=0.
\end{align}
Therefore, the maximal aggregate reward will be obtained, and the convergence rate is $\mathcal{O}(\frac{1}{T})$.

\subsection{With Capacity Unavailable Condition}
In wireless mesh networks, capacity unavailable condition in Algorithm \ref{alg:algorithm1} is used to avoid collisions. 
As analyzed in section \ref{sec:IFRSAlg}, the collision of packet transmissions will lead to an error with the optimal scheduling result.
Define $\delta_c$ and $\delta_p$ are errors between Algorithm \ref{alg:algorithm1} and optimal scheduling without capacity unavailable condition.
Detailed expressions are provided in Appendix E in \cite{DDLArt}, and the main result is shown below.

\textbf{Theorem 3} Without capacity unavailable condition, $\delta_c$ and $\delta_p$ occur, and the optimality gap between the objective and optimal aggregate reward $R^{*}_{r}$ is
\begin{align}\notag
    \frac{1}{T} \sum_{t=0}^{T-1}\sum_{f\in\mathcal{F}}\sum_{\sigma\in \mathcal{P}_f(t)}  w_f \left\{ R^{\pi}_{r}\left[s_{\sigma}(t)\right ]-R^{*}_{r} \right\}\leq \\
    \frac{g(N,F,C,\tau_{\max})+q(\delta_c,\delta_p)}{T}.
\end{align}
where $q$ is the function of the errors shown in Appendix E in \cite{DDLArt}, and it also has a limited value.
It is worth noting that optimality can still be achieved in the presence of errors.

\begin{figure*}[!t]
	\centering
	\subfigure[Throughput under different arrival rates.]{
		\includegraphics[width=.32\linewidth]{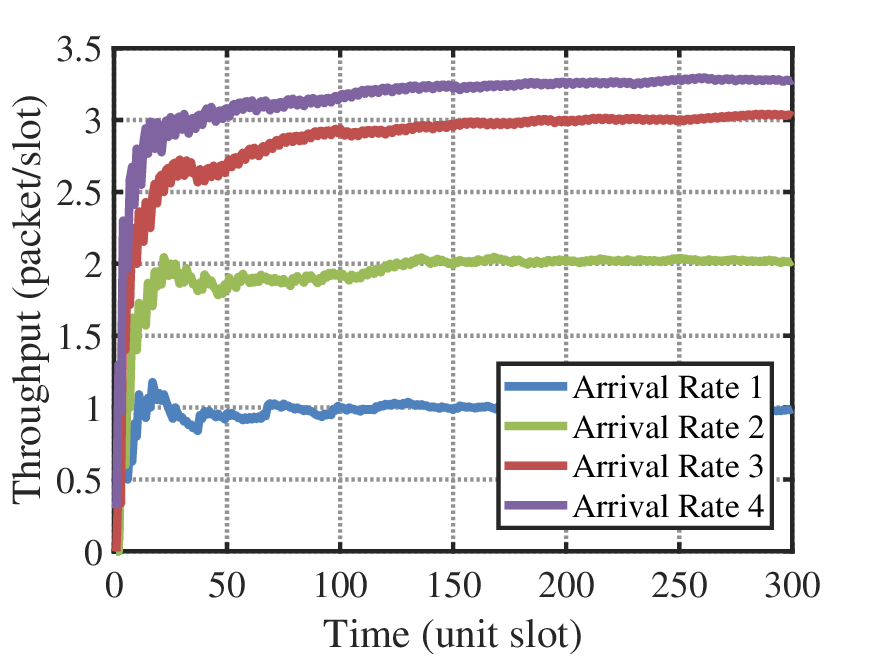}}
	\subfigure[Comparisons without auxiliary reward.]{
		\includegraphics[width=.32\linewidth]{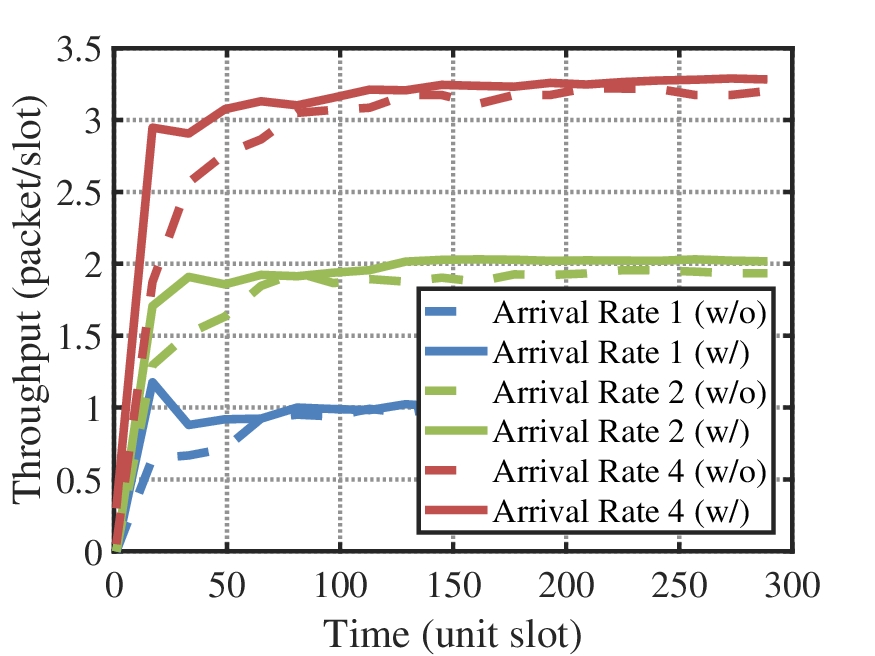}}
    \subfigure[Convergence time of different number of network nodes.]{
		\includegraphics[width=.32\linewidth]{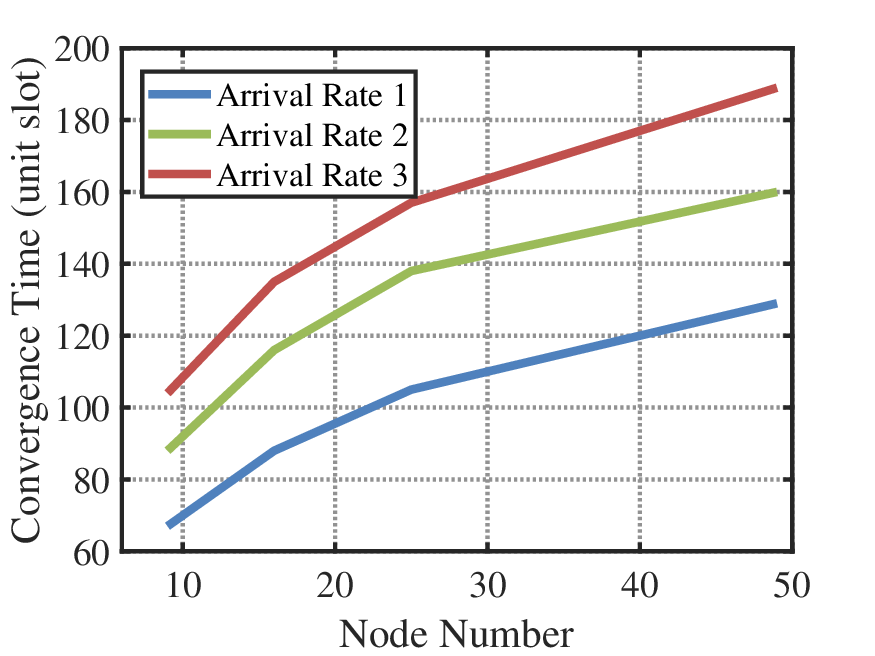}}
	\caption{Verification of optimality and convergence of the PGDS method.}
	\label{fig:NUValid}
\end{figure*}

\begin{figure}[!t]
	\centering
	\subfigure[Tree topology.]{
		\includegraphics[width=.28\linewidth]{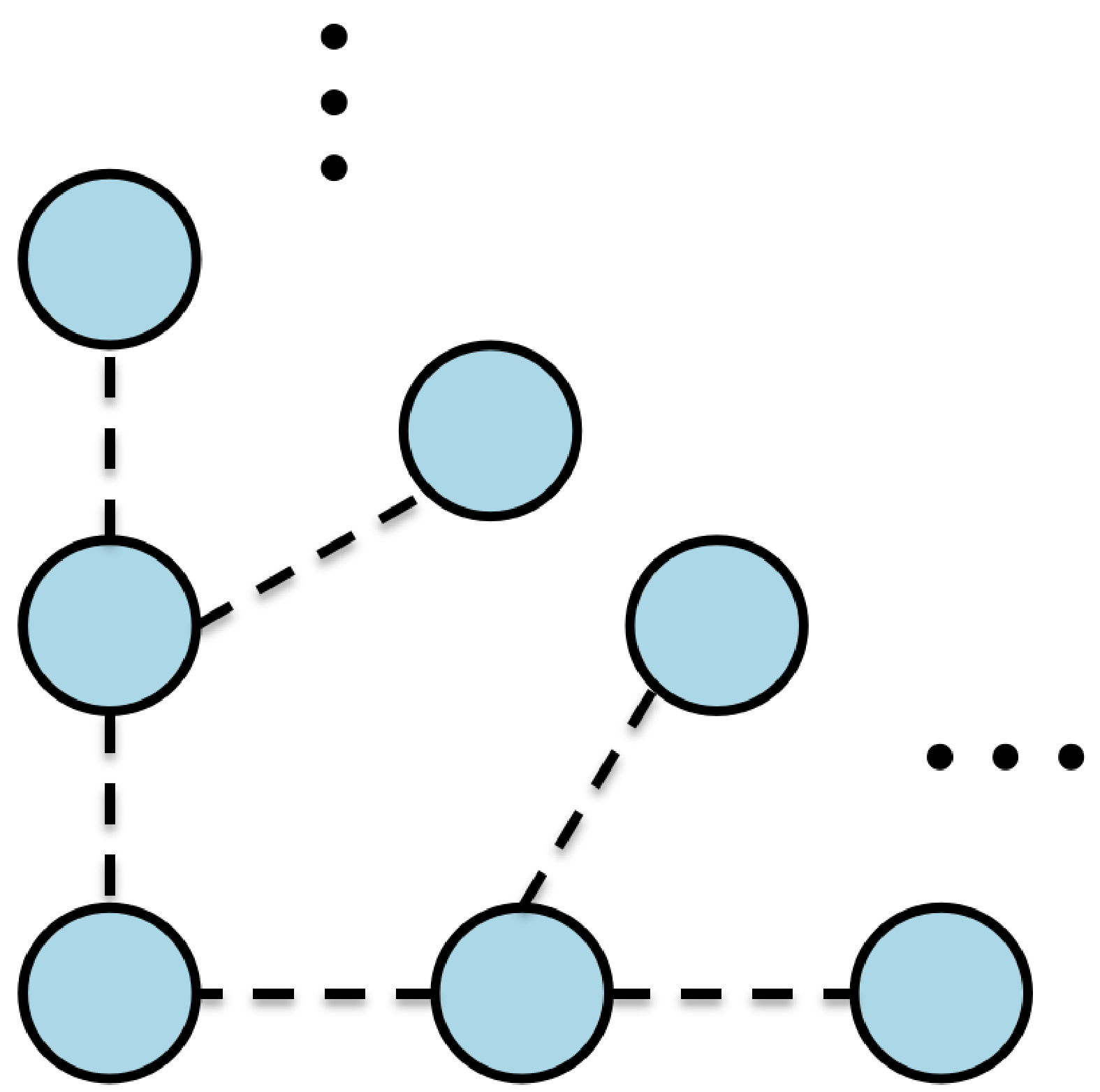}}
	\subfigure[Grid topology.]{
		\includegraphics[width=.3\linewidth]{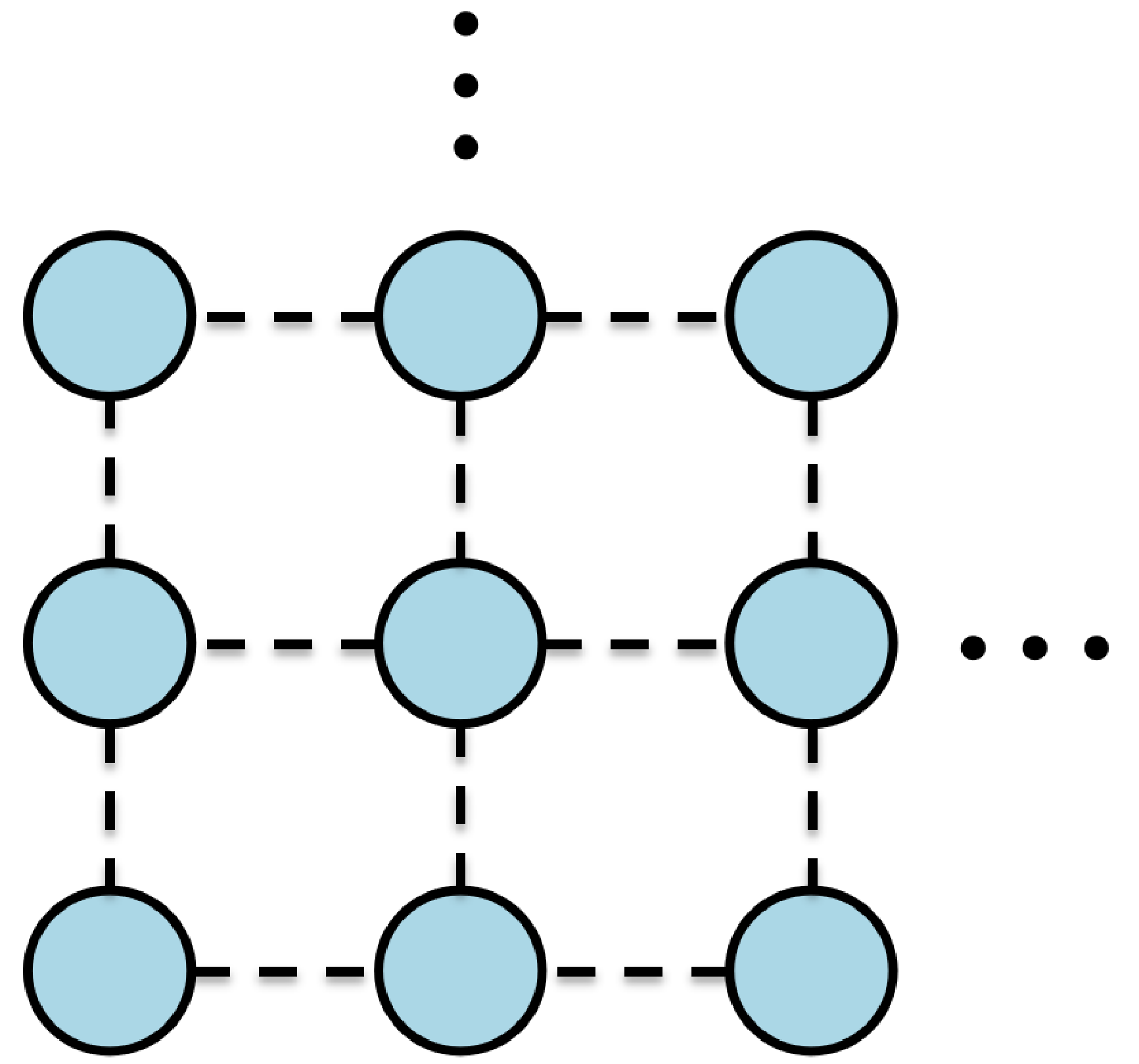}}
    \subfigure[Star topology.]{
		\includegraphics[width=.32\linewidth]{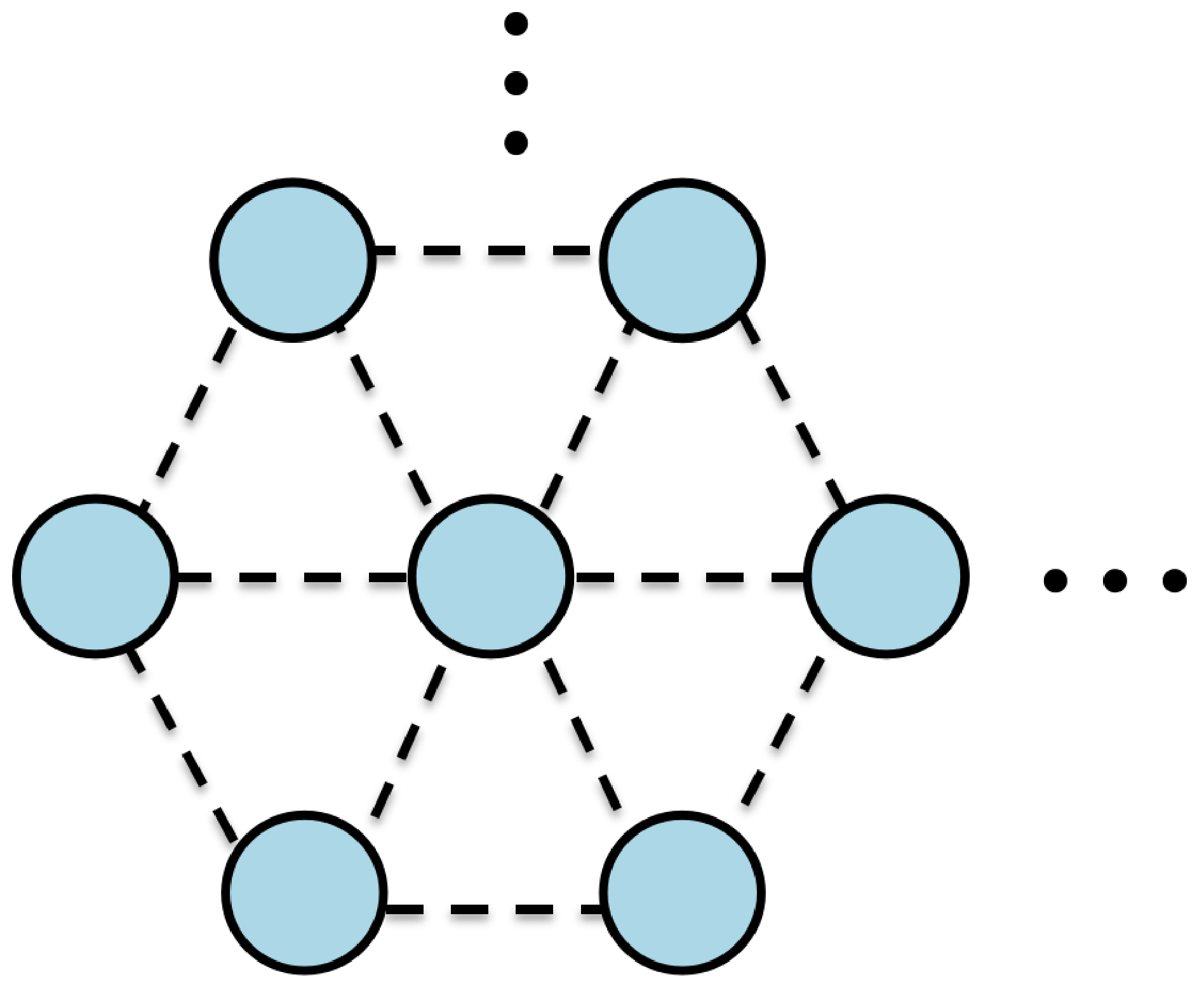}}
	\caption{Three typical topologies of wireless mesh networks.}
	\label{fig:SimSet}
\end{figure}

\section{Performance Evaluations}\label{sec:SimulationResults}
In this section, the optimality and convergence of the PGDS method are first verified, then is to compare the performance difference between this approach and the current best-performing distributed scheduling methods.

\subsection{Verification of Optimality and Convergence}
\begin{figure*}[!t]
	\centering
	\subfigure[Throughput under different topologies.]{
		\includegraphics[width=.32\linewidth]{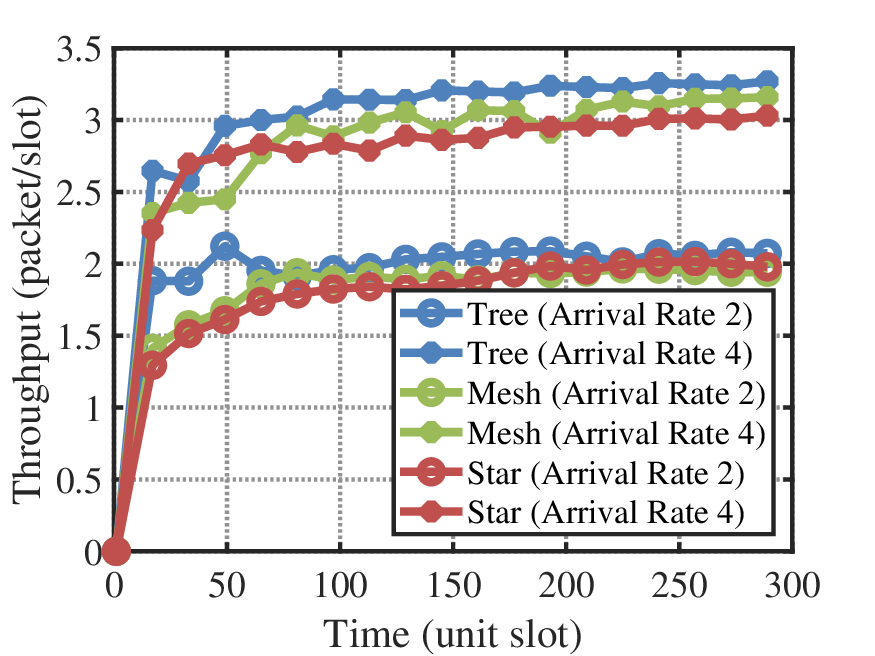}}
	\subfigure[CDF of delay under different topologies.]{
		\includegraphics[width=.32\linewidth]{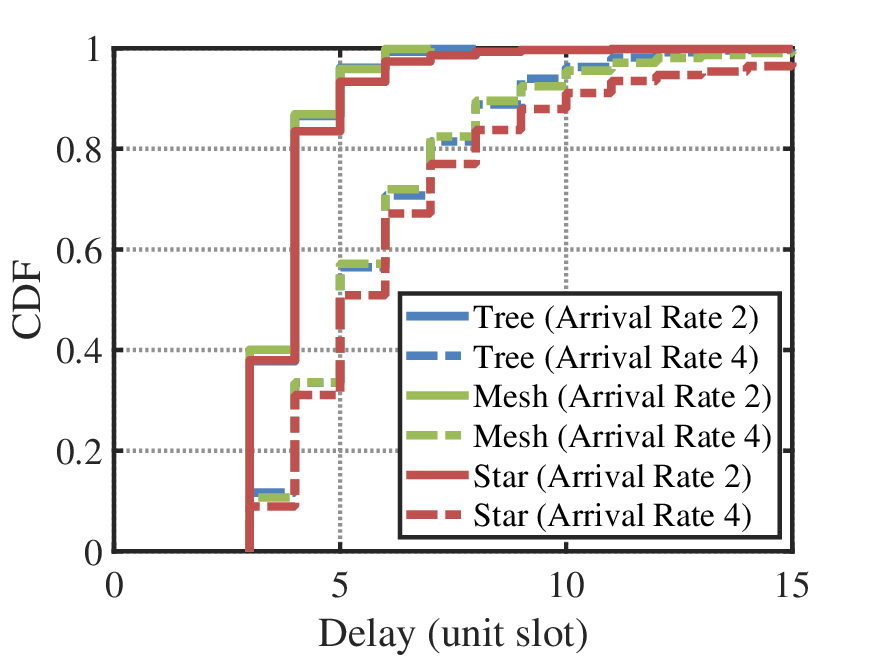}}
    \subfigure[Delay violation probability with different arrival rates.]{
		\includegraphics[width=.32\linewidth]{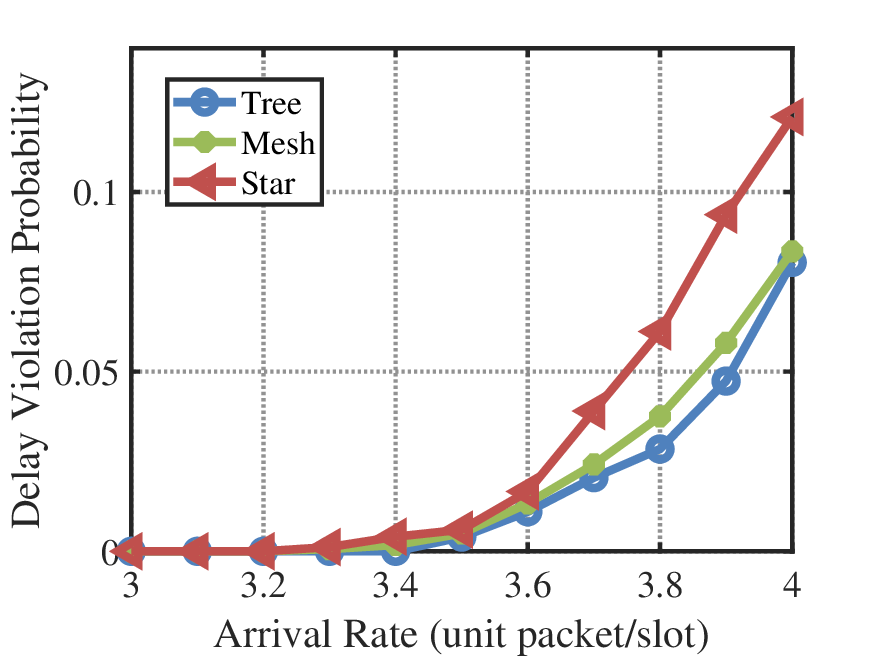}}
	\caption{Throughput and delay performance under different topologies using the PGDS method.}
	\label{fig:TopologyValid}
\end{figure*}

\begin{figure*}[!t]
	\centering
	\subfigure[Throughput comparisons under different arrival rate.]{
		\includegraphics[width=.32\linewidth]{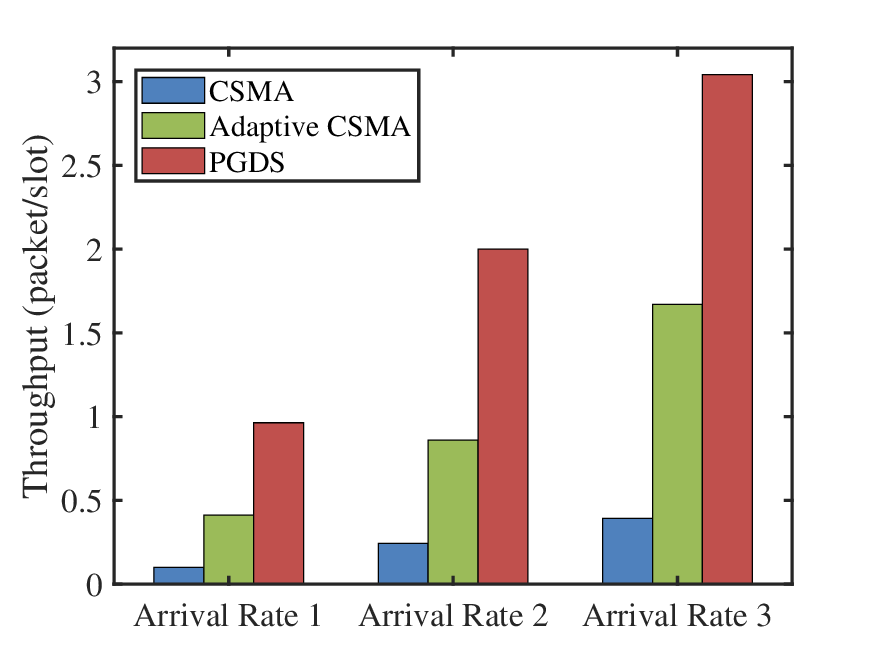}}
	\subfigure[Comparison of delay CDF.]{
		\includegraphics[width=.32\linewidth]{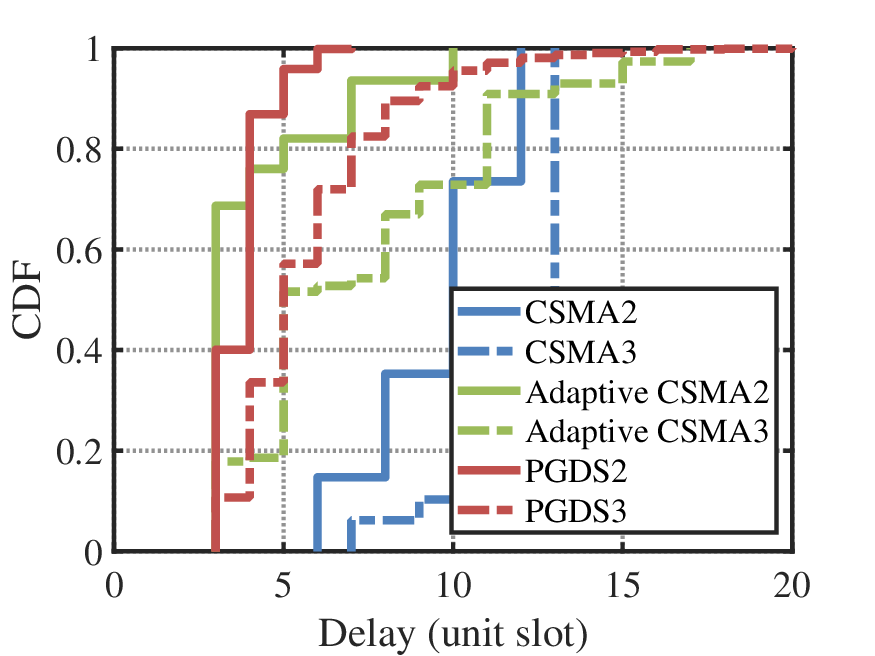}}
    \subfigure[Comparison under network variations.]{
		\includegraphics[width=.32\linewidth]{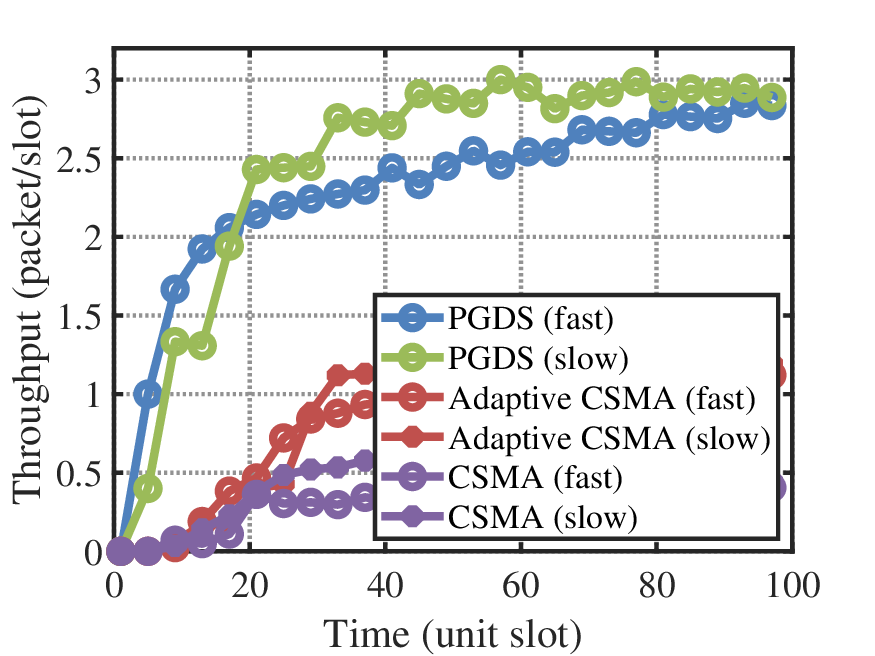}}
	\caption{Comparisons of throughput and delay with existing methods.}
	\label{fig:CompNU}
\end{figure*}

To clearly show the performance of the network, the reward function takes the form of throughput.
Take the mesh topology with 16 nodes and 4 flows as an example.
The relationship between the throughput and different arrival rates is shown in Fig.~\ref{fig:NUValid}(a).
Arrival rates 1,2,3, and 4 represent the sum rate of all the flows are 1, 2, 3, and 4 unit packet/slot, respectively.
Each flow divides all the sum arrival rates equally.
It can be seen that when the arrival rate is small, the convergence rate is faster, and throughput is proportional to the arrival rate.
When the arrival rate is large, throughput tends to be saturated, and is limited by the maximum link capacity, that is, the maximum transmit power per node, the number of subcarriers, and the channel gain.
In addition, the figure can also reflect the rapid convergence, even under saturated network utility, it only takes around 100 units of time slot to converge.

After that, the performance with and without auxiliary reward is shown in Fig~\ref{fig:NUValid}(b).
At different arrival rates, the auxiliary reward can significantly speed up the convergence.
Note that the effect of the auxiliary reward on improving the convergence speed does not increase much as the arrival rate increases.
The main reason is that the auxiliary reward is only effective when the primary reward does not return in the initial stage. 
Regardless of the arrival rate, the effect of the auxiliary reward will continue to decrease as soon as the primary reward starts to generate.

Next, the convergence time under different numbers of nodes in the wireless network is carried out in Fig.~\ref{fig:NUValid}(c).
Flows are generated from the node in the lower left corner and need to be delivered to the destination node in the upper right corner.
Since all the paths of the data flow to the destination node can run through all nodes, it can effectively reflect the influence of the number of network nodes.
It can be seen that when the number of nodes increases, the convergence time increases similarly to the log function form, which shows the correctness of the conclusion in Theorem 2.
Note that the increase in the number of flows is similar to the increase in the arrival rate, so increasing the arrival rate can reflect the impact of increasing the number of data flows to a certain extent.
As the arrival rate increases, the convergence time increases less than the linear increase, in a manner similar to the log function.

\subsection{Performance under Different Topologies}
Tree, grid, and star topologies are classic wireless mesh network topologies, which are considered in the simulation and shown in Fig.~\ref{fig:SimSet}.
The difference between the three topologies lies in the number of neighbors of each node. 
A larger number of neighbors means a larger scheduling space, which will lead to an increase in the convergence time.
A time-slotted system is used, where the unit time slot is set as 0.25 ms and the length of the unit packet is 100 bits.
The total number of subcarriers in the network is 20, the maximum transmit power of each node is 20 dBm.
Links adopt the channel model of the 3GPP standard.
The channel changes for a time greater than twice the maximum end-to-end delay.
Each flow has its own arrival rate.
The maximal deadline for flows to reach their destination is assumed to be 10 unit time slots.
The source and destination nodes of data flow need to cover the edge nodes of the network, so as to reflect the impact of the increase of network size.








The performance of the PGDS method in different topologies is shown in Fig.~\ref{fig:TopologyValid}.
The result of throughput is shown in Fig.~\ref{fig:TopologyValid}(a).
Since the tree topology has the smallest action space, it has a fast convergence speed.
With the help of auxiliary rewards, the action Spaces of grid and star topologies are also reduced a lot, but the interference range becomes larger in these two topologies, which leads to a decrease in the convergence speed.
The cumulative distribution function (CDF) of the delay under the three topologies is shown in Fig.~\ref{fig:TopologyValid}(b).
It can be observed that when the arrival rate increases, more packets will stay in the intermediate nodes due to the limited number of available bandwidth in the interference field, increasing the delay.
After that, the delay violation probability is simulated in Fig.~\ref{fig:TopologyValid}(c), where the horizontal coordinate represents the sum of the arrival rates of all data flows.
It can be seen that when the sum of all data flow arrival rates does not exceed the maximum link capacity, PGDS method can effectively ensure the delay, that is, the delay violation probability is about 0.
As the arrival rate increases, more and more packets are stuck at the source node or intermediate node. The TTD of these packets will continue to decrease until it drops to 0, that is, the packet is invalid.
Therefore, packets are continuously discarded, which leads to an increase in the delay violation probability.
The increase in delay violation probability means that the source node needs to reduce its data generation rate, which belongs to the field of congestion control and is not considered in this paper.


\subsection{Comparisons with Existing Methods}
Finally, the performance of the PGDS method is compared with that of existing methods.
To illustrate the improved performance of the proposed method, the simulation takes traditional CSMA as a baseline and compares it with the most recent best-performing distributed scheduling method, that is adaptive CSMA in~\cite{9377564}.
In the traditional CSMA method, the node listens to the channel at each unit slot, and performs backoff if the channel is idle, so that the counter time is reduced by 1 unit slot, and the data is transmitted when the counter time is 0. 
If a conflict occurs, double the counter time.
In the Adaptive CSMA method, each node has a counter for its connected links, and the time generation of the counter follows the exponential distribution of the mean of a parameter as a unit time slot, and the adjustment of this parameter is in~\cite{9377564}.
The performance comparison of throughput is shown in Fig.~\ref{fig:CompNU}(a), where PGDS method outperforms the existing methods.
The reason is analyzed as follows.
Since CSMA only transmits data according to the parameter of the timer, this parameter does not reflect the performance of transmitting the data flow from source to destination, so the performance is the worst. 
Adaptive CSMA can adjust the timer size through the interaction of local nodes, but cannot obtain end-to-end instructions, so a large number of packets are lost when the delay constraint is very tight, resulting in low throughput.
The CDF of the delay under different arrival rates is shown in Fig.~\ref{fig:CompNU}(b).
In CSMA, a large number of data packets are stuck in intermediate nodes, or fall into some kind of cycle and can not reach the destination node, resulting in a long delay.
Adaptive CSMA can reduce this effect by adjusting the window with local information, to reduce the delay. However, only using local information cannot know the end-to-end transmission situation, so this problem cannot be fundamentally solved.
The proposed PGDS method can find the best transmission link according to the end-to-end feedback information, so it can effectively guarantee the end-to-end delay.
Furthermore, the effect of network variation is compared in Fig.~\ref{fig:CompNU}(c).
In the simulation, the average channel quality is the same for each link.
"fast" means that the duration of the channel quality variation on each link is less than the maximum end-to-end delay, and "slow" means that the channel quality variation on each link changes larger than twice the maximum end-to-end delay.
It can be seen that the rapid change of the channel will slow down the convergence speed, the main reason is that the feedback reward information has a lag, which is no longer applicable to the current channel. 
However, since the statistical characteristics of the channel are not changed, it is still able to converge to the optimal value in the end.
Note that the PGDS method does not consider signaling overhead.
If signaling overhead is 20\%, then the performance gain will decrease 20\%, which will be studied in the future.

\section{Conclusion}\label{sec:Conclusion}
In this paper, a policy gradient-based distributed scheduling (PGDS) method with potential-based reward shaping (PBRS) for wireless networks with end-to-end deadline constraints was proposed. 
The method can achieve maximal throughput for wireless mesh networks and outperform the existing methods.
Theoretically, the optimality and convergence rate of the method were derived, and their relationship to network parameters was analyzed in detail. 
Following the proposed distributed method, an interesting problem for future research is how to apply the method to wireless mobile networks. 
How to integrate other technologies, such as beamforming, to further enhance the network performance is also an important topic to study.

\bibliographystyle{ACM-Reference-Format}
\bibliography{sample-base}


\begin{thebibliography}{42}


\ifx \showCODEN    \undefined \def \showCODEN     #1{\unskip}     \fi
\ifx \showDOI      \undefined \def \showDOI       #1{#1}\fi
\ifx \showISBNx    \undefined \def \showISBNx     #1{\unskip}     \fi
\ifx \showISBNxiii \undefined \def \showISBNxiii  #1{\unskip}     \fi
\ifx \showISSN     \undefined \def \showISSN      #1{\unskip}     \fi
\ifx \showLCCN     \undefined \def \showLCCN      #1{\unskip}     \fi
\ifx \shownote     \undefined \def \shownote      #1{#1}          \fi
\ifx \showarticletitle \undefined \def \showarticletitle #1{#1}   \fi
\ifx \showURL      \undefined \def \showURL       {\relax}        \fi
\providecommand\bibfield[2]{#2}
\providecommand\bibinfo[2]{#2}
\providecommand\natexlab[1]{#1}
\providecommand\showeprint[2][]{arXiv:#2}

\bibitem[Akram and Ugurlu(2023)]%
        {10234442}
\bibfield{author}{\bibinfo{person}{Vahid~Khalilpour Akram} {and}
  \bibinfo{person}{Onur Ugurlu}.} \bibinfo{year}{2023}\natexlab{}.
\newblock \showarticletitle{Detecting the Most Vital Articulation Points in
  Wireless Multi-Hop Networks}.
\newblock \bibinfo{journal}{\emph{IEEE/ACM Transactions on Networking}}
  \bibinfo{volume}{31}, \bibinfo{number}{5} (\bibinfo{year}{2023}),
  \bibinfo{pages}{2389--2402}.
\newblock


\bibitem[Altman(1999)]%
        {AltemanCMDP}
\bibfield{author}{\bibinfo{person}{Eitan Altman}.}
  \bibinfo{year}{1999}\natexlab{}.
\newblock \bibinfo{booktitle}{\emph{Constrained Markov Decision Processes}
  (\bibinfo{edition}{1st} ed.)}.
\newblock \bibinfo{publisher}{Routledge}, \bibinfo{address}{New York, USA}.
\newblock
\showISBNx{9781315140223}


\bibitem[Chao and Hsu(2023)]%
        {9678001}
\bibfield{author}{\bibinfo{person}{I-Fen Chao} {and} \bibinfo{person}{Wei-Sheng
  Hsu}.} \bibinfo{year}{2023}\natexlab{}.
\newblock \showarticletitle{A MAC Protocol Design for Maximizing End-to-End
  Throughput and Fairness Guarantee in Chain-Based Multi-Hop Wireless Backhaul
  Networks}.
\newblock \bibinfo{journal}{\emph{IEEE Transactions on Mobile Computing}}
  \bibinfo{volume}{22}, \bibinfo{number}{7} (\bibinfo{year}{2023}),
  \bibinfo{pages}{3743--3756}.
\newblock


\bibitem[Chen and Huang(2018)]%
        {8476232}
\bibfield{author}{\bibinfo{person}{Kun Chen} {and} \bibinfo{person}{Longbo
  Huang}.} \bibinfo{year}{2018}\natexlab{}.
\newblock \showarticletitle{Timely-Throughput Optimal Scheduling With
  Prediction}.
\newblock \bibinfo{journal}{\emph{IEEE/ACM Transactions on Networking}}
  \bibinfo{volume}{26}, \bibinfo{number}{6} (\bibinfo{year}{2018}),
  \bibinfo{pages}{2457--2470}.
\newblock


\bibitem[Deng et~al\mbox{.}(2019)]%
        {8726331}
\bibfield{author}{\bibinfo{person}{Han Deng}, \bibinfo{person}{Tao Zhao}, {and}
  \bibinfo{person}{I-Hong Hou}.} \bibinfo{year}{2019}\natexlab{}.
\newblock \showarticletitle{Online Routing and Scheduling With Capacity
  Redundancy for Timely Delivery Guarantees in Multihop Networks}.
\newblock \bibinfo{journal}{\emph{IEEE/ACM Transactions on Networking}}
  \bibinfo{volume}{27}, \bibinfo{number}{3} (\bibinfo{year}{2019}),
  \bibinfo{pages}{1258--1271}.
\newblock


\bibitem[Ding et~al\mbox{.}(2020)]%
        {NEURIPS2020_5f7695de}
\bibfield{author}{\bibinfo{person}{Dongsheng Ding}, \bibinfo{person}{Kaiqing
  Zhang}, \bibinfo{person}{Tamer Basar}, {and} \bibinfo{person}{Mihailo
  Jovanovic}.} \bibinfo{year}{2020}\natexlab{}.
\newblock \showarticletitle{Natural Policy Gradient Primal-Dual Method for
  Constrained Markov Decision Processes}. In \bibinfo{booktitle}{\emph{Advances
  in Neural Information Processing Systems}}, Vol.~\bibinfo{volume}{33}.
  \bibinfo{publisher}{Curran Associates, Inc.}, \bibinfo{pages}{8378--8390}.
\newblock


\bibitem[Hoteit et~al\mbox{.}(2013)]%
        {6265054}
\bibfield{author}{\bibinfo{person}{Sahar Hoteit}, \bibinfo{person}{Stefano
  Secci}, \bibinfo{person}{Rami Langar}, {and} \bibinfo{person}{Guy Pujolle}.}
  \bibinfo{year}{2013}\natexlab{}.
\newblock \showarticletitle{A Nucleolus-Based Approach for Resource Allocation
  in OFDMA Wireless Mesh Networks}.
\newblock \bibinfo{journal}{\emph{IEEE Transactions on Mobile Computing}}
  \bibinfo{volume}{12}, \bibinfo{number}{11} (\bibinfo{year}{2013}),
  \bibinfo{pages}{2145--2154}.
\newblock


\bibitem[Hou and Kumar(2010a)]%
        {5462070}
\bibfield{author}{\bibinfo{person}{I-Hong Hou} {and} \bibinfo{person}{P.~R.
  Kumar}.} \bibinfo{year}{2010}\natexlab{a}.
\newblock \showarticletitle{Utility Maximization for Delay Constrained QoS in
  Wireless}. In \bibinfo{booktitle}{\emph{2010 Proceedings IEEE INFOCOM}}.
  \bibinfo{pages}{1--9}.
\newblock


\bibitem[Hou and Kumar(2010b)]%
        {MobiHocSignleHop}
\bibfield{author}{\bibinfo{person}{I-Hong Hou} {and} \bibinfo{person}{P.~R.
  Kumar}.} \bibinfo{year}{2010}\natexlab{b}.
\newblock \showarticletitle{Utility-optimal scheduling in time-varying wireless
  networks with delay constraints}. In \bibinfo{booktitle}{\emph{Proceedings of
  the Eleventh ACM International Symposium on Mobile Ad Hoc Networking and
  Computing}} (Chicago, Illinois, USA) \emph{(\bibinfo{series}{MobiHoc '10})}.
  \bibinfo{publisher}{Association for Computing Machinery},
  \bibinfo{address}{New York, NY, USA}, \bibinfo{pages}{31–40}.
\newblock
\showISBNx{9781450301831}


\bibitem[Hsieh and Sivakumar(2001)]%
        {SIGMETRICS01}
\bibfield{author}{\bibinfo{person}{Hung-Yun Hsieh} {and}
  \bibinfo{person}{Raghupathy Sivakumar}.} \bibinfo{year}{2001}\natexlab{}.
\newblock \showarticletitle{Performance comparison of cellular and multi-hop
  wireless networks: a quantitative study}. In
  \bibinfo{booktitle}{\emph{Proceedings of the 2001 ACM SIGMETRICS
  International Conference on Measurement and Modeling of Computer Systems}}
  (Cambridge, Massachusetts, USA) \emph{(\bibinfo{series}{SIGMETRICS '01})}.
  \bibinfo{publisher}{Association for Computing Machinery},
  \bibinfo{address}{New York, NY, USA}, \bibinfo{pages}{113–122}.
\newblock
\showISBNx{1581133340}


\bibitem[Huang et~al\mbox{.}(2022)]%
        {9705511}
\bibfield{author}{\bibinfo{person}{Cheng Huang}, \bibinfo{person}{Xin Wang},
  {and} \bibinfo{person}{Xudong Wang}.} \bibinfo{year}{2022}\natexlab{}.
\newblock \showarticletitle{Effective-Capacity-Based Resource Allocation for
  End-to-End Multi-Connectivity in 5G IAB Networks}.
\newblock \bibinfo{journal}{\emph{IEEE Transactions on Wireless
  Communications}} \bibinfo{volume}{21}, \bibinfo{number}{8}
  (\bibinfo{year}{2022}), \bibinfo{pages}{6302--6316}.
\newblock


\bibitem[Jahromizadeh and Rakocevic(2014)]%
        {6619407}
\bibfield{author}{\bibinfo{person}{Soroush Jahromizadeh} {and}
  \bibinfo{person}{Veselin Rakocevic}.} \bibinfo{year}{2014}\natexlab{}.
\newblock \showarticletitle{Joint Rate Control and Scheduling for Providing
  Bounded Delay With High Efficiency in Multihop Wireless Networks}.
\newblock \bibinfo{journal}{\emph{IEEE/ACM Transactions on Networking}}
  \bibinfo{volume}{22}, \bibinfo{number}{5} (\bibinfo{year}{2014}),
  \bibinfo{pages}{1686--1698}.
\newblock


\bibitem[Jang et~al\mbox{.}(2014)]%
        {6847949}
\bibfield{author}{\bibinfo{person}{Hyeryung Jang}, \bibinfo{person}{Se-Young
  Yun}, \bibinfo{person}{Jinwoo Shin}, {and} \bibinfo{person}{Yung Yi}.}
  \bibinfo{year}{2014}\natexlab{}.
\newblock \showarticletitle{Distributed learning for utility maximization over
  CSMA-based wireless multihop networks}. In \bibinfo{booktitle}{\emph{IEEE
  INFOCOM 2014 - IEEE Conference on Computer Communications}}.
  \bibinfo{pages}{280--288}.
\newblock


\bibitem[Ji et~al\mbox{.}(2023)]%
        {10125013}
\bibfield{author}{\bibinfo{person}{Jiequ Ji}, \bibinfo{person}{Xiangyu Ren},
  \bibinfo{person}{Lin Cai}, {and} \bibinfo{person}{Kun Zhu}.}
  \bibinfo{year}{2023}\natexlab{}.
\newblock \showarticletitle{Downlink Scheduler for Delay Guaranteed Services
  Using Deep Reinforcement Learning}.
\newblock \bibinfo{journal}{\emph{IEEE Transactions on Mobile Computing}}
  (\bibinfo{year}{2023}), \bibinfo{pages}{1--14}.
\newblock


\bibitem[Jiang and Walrand(2010)]%
        {5340575}
\bibfield{author}{\bibinfo{person}{Libin Jiang} {and} \bibinfo{person}{Jean
  Walrand}.} \bibinfo{year}{2010}\natexlab{}.
\newblock \showarticletitle{A Distributed CSMA Algorithm for Throughput and
  Utility Maximization in Wireless Networks}.
\newblock \bibinfo{journal}{\emph{IEEE/ACM Transactions on Networking}}
  \bibinfo{volume}{18}, \bibinfo{number}{3} (\bibinfo{year}{2010}),
  \bibinfo{pages}{960--972}.
\newblock


\bibitem[Jung and Lee(2021)]%
        {9172113}
\bibfield{author}{\bibinfo{person}{Ji-Young Jung} {and}
  \bibinfo{person}{Jung-Ryun Lee}.} \bibinfo{year}{2021}\natexlab{}.
\newblock \showarticletitle{Distributed Fair Resource Allocation Algorithm in
  Wireless Multihop Network Systems}.
\newblock \bibinfo{journal}{\emph{IEEE Systems Journal}} \bibinfo{volume}{15},
  \bibinfo{number}{4} (\bibinfo{year}{2021}), \bibinfo{pages}{4713--4724}.
\newblock


\bibitem[Kadota et~al\mbox{.}(2018)]%
        {8486307}
\bibfield{author}{\bibinfo{person}{Igor Kadota}, \bibinfo{person}{Abhishek
  Sinha}, {and} \bibinfo{person}{Eytan Modiano}.}
  \bibinfo{year}{2018}\natexlab{}.
\newblock \showarticletitle{Optimizing Age of Information in Wireless Networks
  with Throughput Constraints}. In \bibinfo{booktitle}{\emph{IEEE INFOCOM 2018
  - IEEE Conference on Computer Communications}}. \bibinfo{pages}{1844--1852}.
\newblock


\bibitem[Li and Eryilmaz(2012)]%
        {6155625}
\bibfield{author}{\bibinfo{person}{Ruogu Li} {and} \bibinfo{person}{Atilla
  Eryilmaz}.} \bibinfo{year}{2012}\natexlab{}.
\newblock \showarticletitle{Scheduling for End-to-End Deadline-Constrained
  Traffic With Reliability Requirements in Multihop Networks}.
\newblock \bibinfo{journal}{\emph{IEEE/ACM Transactions on Networking}}
  \bibinfo{volume}{20}, \bibinfo{number}{5} (\bibinfo{year}{2012}),
  \bibinfo{pages}{1649--1662}.
\newblock


\bibitem[Liu et~al\mbox{.}(2023)]%
        {10159148}
\bibfield{author}{\bibinfo{person}{Kang Liu}, \bibinfo{person}{Wei Quan},
  \bibinfo{person}{Nan Cheng}, \bibinfo{person}{Xue Zhang},
  \bibinfo{person}{Liang Guo}, \bibinfo{person}{Deyun Gao}, {and}
  \bibinfo{person}{Hongke Zhang}.} \bibinfo{year}{2023}\natexlab{}.
\newblock \showarticletitle{Deadline-Constrained Multi-Agent Collaborative
  Transmission for Delay-Sensitive Applications}.
\newblock \bibinfo{journal}{\emph{IEEE Transactions on Cognitive Communications
  and Networking}} \bibinfo{volume}{9}, \bibinfo{number}{5}
  (\bibinfo{year}{2023}), \bibinfo{pages}{1370--1384}.
\newblock


\bibitem[Liu et~al\mbox{.}(2020)]%
        {9144447}
\bibfield{author}{\bibinfo{person}{Qingyu Liu}, \bibinfo{person}{Haibo Zeng},
  {and} \bibinfo{person}{Minghua Chen}.} \bibinfo{year}{2020}\natexlab{}.
\newblock \showarticletitle{Network Utility Maximization Under Maximum Delay
  Constraints and Throughput Requirements}.
\newblock \bibinfo{journal}{\emph{IEEE/ACM Transactions on Networking}}
  \bibinfo{volume}{28}, \bibinfo{number}{5} (\bibinfo{year}{2020}),
  \bibinfo{pages}{2132--2145}.
\newblock


\bibitem[Liu et~al\mbox{.}(2019)]%
        {LIU2019102007}
\bibfield{author}{\bibinfo{person}{Xin Liu}, \bibinfo{person}{Weichang Wang},
  {and} \bibinfo{person}{Lei Ying}.} \bibinfo{year}{2019}\natexlab{}.
\newblock \showarticletitle{Spatial–temporal routing for supporting
  end-to-end hard deadlines in multi-hop networks}.
\newblock \bibinfo{journal}{\emph{Performance Evaluation}}
  \bibinfo{volume}{135} (\bibinfo{year}{2019}), \bibinfo{pages}{102007}.
\newblock
\showISSN{0166-5316}


\bibitem[Lou et~al\mbox{.}(2023)]%
        {9792409}
\bibfield{author}{\bibinfo{person}{Jiadong Lou}, \bibinfo{person}{Xu Yuan},
  \bibinfo{person}{Purushottam Sigdel}, \bibinfo{person}{Xiaoqi Qin},
  \bibinfo{person}{Sastry Kompella}, {and} \bibinfo{person}{Nian-Feng Tzeng}.}
  \bibinfo{year}{2023}\natexlab{}.
\newblock \showarticletitle{Age of Information Optimization in Multi-Channel
  Based Multi-Hop Wireless Networks}.
\newblock \bibinfo{journal}{\emph{IEEE Transactions on Mobile Computing}}
  \bibinfo{volume}{22}, \bibinfo{number}{10} (\bibinfo{year}{2023}),
  \bibinfo{pages}{5719--5732}.
\newblock


\bibitem[Neely(2011)]%
        {5934971}
\bibfield{author}{\bibinfo{person}{Michael~J. Neely}.}
  \bibinfo{year}{2011}\natexlab{}.
\newblock \showarticletitle{Opportunistic scheduling with worst case delay
  guarantees in single and multi-hop networks}. In
  \bibinfo{booktitle}{\emph{2011 Proceedings IEEE INFOCOM}}.
  \bibinfo{pages}{1728--1736}.
\newblock


\bibitem[Neely(2013)]%
        {6180023}
\bibfield{author}{\bibinfo{person}{Michael~J. Neely}.}
  \bibinfo{year}{2013}\natexlab{}.
\newblock \showarticletitle{Delay-Based Network Utility Maximization}.
\newblock \bibinfo{journal}{\emph{IEEE/ACM Transactions on Networking}}
  \bibinfo{volume}{21}, \bibinfo{number}{1} (\bibinfo{year}{2013}),
  \bibinfo{pages}{41--54}.
\newblock


\bibitem[Ng et~al\mbox{.}(1999)]%
        {RewardShaping}
\bibfield{author}{\bibinfo{person}{Andrew~Y Ng}, \bibinfo{person}{Daishi
  Harada}, {and} \bibinfo{person}{Stuart Russell}.}
  \bibinfo{year}{1999}\natexlab{}.
\newblock \showarticletitle{Policy invariance under reward transformations:
  Theory and application to reward shaping}. In
  \bibinfo{booktitle}{\emph{Proceedings of the Twentieth International
  Conference on International Conference on Machine Learning}} (Washington, DC,
  USA) \emph{(\bibinfo{series}{ICML'99})}. \bibinfo{publisher}{AAAI Press},
  \bibinfo{pages}{278–287}.
\newblock


\bibitem[Ni et~al\mbox{.}(2012)]%
        {2177101}
\bibfield{author}{\bibinfo{person}{Jian Ni}, \bibinfo{person}{Bo Tan}, {and}
  \bibinfo{person}{R. Srikant}.} \bibinfo{year}{2012}\natexlab{}.
\newblock \showarticletitle{Q-CSMA: Queue-Length-Based CSMA/CA Algorithms for
  Achieving Maximum Throughput and Low Delay in Wireless Networks}.
\newblock \bibinfo{journal}{\emph{IEEE/ACM Transactions on Networking}}
  \bibinfo{volume}{20}, \bibinfo{number}{3} (\bibinfo{date}{jun}
  \bibinfo{year}{2012}), \bibinfo{pages}{825–836}.
\newblock
\showISSN{1063-6692}


\bibitem[Pagin et~al\mbox{.}(2022)]%
        {9500058}
\bibfield{author}{\bibinfo{person}{Matteo Pagin}, \bibinfo{person}{Tommaso
  Zugno}, \bibinfo{person}{Michele Polese}, {and} \bibinfo{person}{Michele
  Zorzi}.} \bibinfo{year}{2022}\natexlab{}.
\newblock \showarticletitle{Resource Management for 5G NR Integrated Access and
  Backhaul: A Semi-Centralized Approach}.
\newblock \bibinfo{journal}{\emph{IEEE Transactions on Wireless
  Communications}} \bibinfo{volume}{21}, \bibinfo{number}{2}
  (\bibinfo{year}{2022}), \bibinfo{pages}{753--767}.
\newblock


\bibitem[Palazzi(2018)]%
        {MobiHoc18Multihop}
\bibfield{author}{\bibinfo{person}{Claudio~E. Palazzi}.}
  \bibinfo{year}{2018}\natexlab{}.
\newblock \showarticletitle{High Bandwidth and Low Delay over Wireless Multihop
  Networks} \emph{(\bibinfo{series}{Mobihoc '18})}.
  \bibinfo{publisher}{Association for Computing Machinery},
  \bibinfo{address}{New York, NY, USA}, \bibinfo{pages}{306–307}.
\newblock
\showISBNx{9781450357708}


\bibitem[Poor(2000)]%
        {Gradient}
\bibfield{author}{\bibinfo{person}{R.~D. Poor}.}
  \bibinfo{year}{2000}\natexlab{}.
\newblock \showarticletitle{Gradient Routing in Ad Hoc Networks}.
\newblock \bibinfo{journal}{\emph{Massachusetts Institute of Technology}}
  (\bibinfo{year}{2000}).
\newblock


\bibitem[Satija et~al\mbox{.}(2020)]%
        {ValueBasedICML}
\bibfield{author}{\bibinfo{person}{Harsh Satija}, \bibinfo{person}{Philip
  Amortila}, {and} \bibinfo{person}{Joelle Pineau}.}
  \bibinfo{year}{2020}\natexlab{}.
\newblock \showarticletitle{Constrained {M}arkov Decision Processes via
  Backward Value Functions}. In \bibinfo{booktitle}{\emph{Proceedings of the
  37th International Conference on Machine Learning}}
  \emph{(\bibinfo{series}{Proceedings of Machine Learning Research},
  Vol.~\bibinfo{volume}{119})}. \bibinfo{publisher}{PMLR},
  \bibinfo{pages}{8502--8511}.
\newblock


\bibitem[Singh and Kumar(2019)]%
        {8485769}
\bibfield{author}{\bibinfo{person}{Rahul Singh} {and} \bibinfo{person}{P.~R.
  Kumar}.} \bibinfo{year}{2019}\natexlab{}.
\newblock \showarticletitle{Throughput Optimal Decentralized Scheduling of
  Multihop Networks With End-to-End Deadline Constraints: Unreliable Links}.
\newblock \bibinfo{journal}{\emph{IEEE Trans. Automat. Control}}
  \bibinfo{volume}{64}, \bibinfo{number}{1} (\bibinfo{year}{2019}),
  \bibinfo{pages}{127--142}.
\newblock


\bibitem[Singh and Kumar(2021)]%
        {9377564}
\bibfield{author}{\bibinfo{person}{Rahul Singh} {and} \bibinfo{person}{P.~R.
  Kumar}.} \bibinfo{year}{2021}\natexlab{}.
\newblock \showarticletitle{Adaptive CSMA for Decentralized Scheduling of
  Multi-Hop Networks With End-to-End Deadline Constraints}.
\newblock \bibinfo{journal}{\emph{IEEE/ACM Transactions on Networking}}
  \bibinfo{volume}{29}, \bibinfo{number}{3} (\bibinfo{year}{2021}),
  \bibinfo{pages}{1224--1237}.
\newblock


\bibitem[Swamy et~al\mbox{.}(2017)]%
        {7878682}
\bibfield{author}{\bibinfo{person}{Peruru~Subrahmanya Swamy},
  \bibinfo{person}{Radha~Krishna Ganti}, {and} \bibinfo{person}{Krishna
  Jagannathan}.} \bibinfo{year}{2017}\natexlab{}.
\newblock \showarticletitle{Adaptive CSMA Under the SINR Model: Efficient
  Approximation Algorithms for Throughput and Utility Maximization}.
\newblock \bibinfo{journal}{\emph{IEEE/ACM Transactions on Networking}}
  \bibinfo{volume}{25}, \bibinfo{number}{4} (\bibinfo{year}{2017}),
  \bibinfo{pages}{1968--1981}.
\newblock


\bibitem[Tripathi et~al\mbox{.}(2023)]%
        {9870687}
\bibfield{author}{\bibinfo{person}{Vishrant Tripathi}, \bibinfo{person}{Rajat
  Talak}, {and} \bibinfo{person}{Eytan Modiano}.}
  \bibinfo{year}{2023}\natexlab{}.
\newblock \showarticletitle{Information Freshness in Multihop Wireless
  Networks}.
\newblock \bibinfo{journal}{\emph{IEEE/ACM Transactions on Networking}}
  \bibinfo{volume}{31}, \bibinfo{number}{2} (\bibinfo{year}{2023}),
  \bibinfo{pages}{784--799}.
\newblock


\bibitem[Tsanikidis and Ghaderi(2022)]%
        {MoboHoc22Scheduling}
\bibfield{author}{\bibinfo{person}{Christos Tsanikidis} {and}
  \bibinfo{person}{Javad Ghaderi}.} \bibinfo{year}{2022}\natexlab{}.
\newblock \showarticletitle{Online Scheduling and Routing with End-to-End
  Deadline Constraints in Multihop Wireless Networks}
  \emph{(\bibinfo{series}{MobiHoc '22})}. \bibinfo{publisher}{Association for
  Computing Machinery}, \bibinfo{address}{New York, NY, USA},
  \bibinfo{pages}{11–20}.
\newblock
\showISBNx{9781450391658}


\bibitem[Tsanikidis and Ghaderi(2023)]%
        {3626781}
\bibfield{author}{\bibinfo{person}{Christos Tsanikidis} {and}
  \bibinfo{person}{Javad Ghaderi}.} \bibinfo{year}{2023}\natexlab{}.
\newblock \showarticletitle{Near-Optimal Packet Scheduling in Multihop Networks
  with End-to-End Deadline Constraints}.
\newblock  \bibinfo{volume}{7}, \bibinfo{number}{3}, Article
  \bibinfo{articleno}{50} (\bibinfo{date}{dec} \bibinfo{year}{2023}),
  \bibinfo{numpages}{32}~pages.
\newblock


\bibitem[Wiewiora et~al\mbox{.}(2003)]%
        {3041938}
\bibfield{author}{\bibinfo{person}{Eric Wiewiora}, \bibinfo{person}{Garrison
  Cottrell}, {and} \bibinfo{person}{Charles Elkan}.}
  \bibinfo{year}{2003}\natexlab{}.
\newblock \showarticletitle{Principled Methods for Advising Reinforcement
  Learning Agents}. In \bibinfo{booktitle}{\emph{Proceedings of the Twentieth
  International Conference on International Conference on Machine Learning}}
  (Washington, DC, USA) \emph{(\bibinfo{series}{ICML'03})}.
  \bibinfo{publisher}{AAAI Press}, \bibinfo{pages}{792–799}.
\newblock
\showISBNx{1577351894}


\bibitem[Xin and Wang({[n.\,d.]})]%
        {DDLArt}
\bibfield{author}{\bibinfo{person}{Wang Xin} {and} \bibinfo{person}{Xudong
  Wang}.} \bibinfo{year}{[n.\,d.]}\natexlab{}.
\newblock \showarticletitle{Distributed Scheduling for Throughput Maximization
  under Deadline Constraint in Wireless Mesh Networks}.
\newblock \bibinfo{journal}{\emph{accessible at
  \url{https://wanglab.sjtu.edu.cn/userfiles/files/Article.pdf}}}
  (\bibinfo{year}{[n.\,d.]}).
\newblock


\bibitem[Xiong et~al\mbox{.}(2011)]%
        {5753559}
\bibfield{author}{\bibinfo{person}{Haozhi Xiong}, \bibinfo{person}{Ruogu Li},
  \bibinfo{person}{Atilla Eryilmaz}, {and} \bibinfo{person}{Eylem Ekici}.}
  \bibinfo{year}{2011}\natexlab{}.
\newblock \showarticletitle{Delay-Aware Cross-Layer Design for Network Utility
  Maximization in Multi-Hop Networks}.
\newblock \bibinfo{journal}{\emph{IEEE Journal on Selected Areas in
  Communications}} \bibinfo{volume}{29}, \bibinfo{number}{5}
  (\bibinfo{year}{2011}), \bibinfo{pages}{951--959}.
\newblock


\bibitem[Xue and Ekici(2013)]%
        {6389738}
\bibfield{author}{\bibinfo{person}{Dongyue Xue} {and} \bibinfo{person}{Eylem
  Ekici}.} \bibinfo{year}{2013}\natexlab{}.
\newblock \showarticletitle{Delay-Guaranteed Cross-Layer Scheduling in Multihop
  Wireless Networks}.
\newblock \bibinfo{journal}{\emph{IEEE/ACM Transactions on Networking}}
  \bibinfo{volume}{21}, \bibinfo{number}{6} (\bibinfo{year}{2013}),
  \bibinfo{pages}{1696--1707}.
\newblock


\bibitem[Zhao and Lin(2016)]%
        {7112191}
\bibfield{author}{\bibinfo{person}{Shizhen Zhao} {and} \bibinfo{person}{Xiaojun
  Lin}.} \bibinfo{year}{2016}\natexlab{}.
\newblock \showarticletitle{Design of Scheduling Algorithms for End-to-End
  Backlog Minimization in Wireless Multi-Hop Networks Under $K$-Hop
  Interference Models}.
\newblock \bibinfo{journal}{\emph{IEEE/ACM Transactions on Networking}}
  \bibinfo{volume}{24}, \bibinfo{number}{2} (\bibinfo{year}{2016}),
  \bibinfo{pages}{1265--1278}.
\newblock


\bibitem[Zhou et~al\mbox{.}(2012)]%
        {6166337}
\bibfield{author}{\bibinfo{person}{Anfu Zhou}, \bibinfo{person}{Min Liu},
  \bibinfo{person}{Zhongcheng Li}, {and} \bibinfo{person}{Eryk Dutkiewicz}.}
  \bibinfo{year}{2012}\natexlab{}.
\newblock \showarticletitle{Cross-Layer Design for Proportional Delay
  Differentiation and Network Utility Maximization in Multi-Hop Wireless
  Networks}.
\newblock \bibinfo{journal}{\emph{IEEE Transactions on Wireless
  Communications}} \bibinfo{volume}{11}, \bibinfo{number}{4}
  (\bibinfo{year}{2012}), \bibinfo{pages}{1446--1455}.
\newblock


\end{thebibliography}

\newpage

\appendix 

\section{Problem decomposition} \label{apdx:PD}
Recall the problem formulation in \eqref{eq:MIOP}, the Lagrangian decomposition is denoted as
\begin{align}\notag
	\mathfrak{L}(\pi,\lambda,\mu) &= \lim_{T\to \infty} \frac{1}{T} \sum_{t=0}^{T-1}\sum_{f\in\mathcal{F}}\sum_{\sigma\in \mathcal{P}_f(t)}w_fR_{r}^{\pi}\left[s_{\sigma}(t)\right]-\\ \notag
	&\sum_{i\in\mathcal{V}}\lambda_i\frac{1}{T} \sum_{t=0}^{T-1}\sum_{f\in\mathcal{F}}\sum_{\sigma\in \mathcal{P}_{fi}(t)\cup \mathcal{P}_{\mathcal{I}_i}(t)}D_{\rm c}^{\pi}\left[s_{\sigma}(t)\right]-\\ \notag
	&\sum_{i\in\mathcal{V}}\mu_{i} \frac{1}{T} \sum_{t=0}^{T-1}\sum_{f\in\mathcal{F}}\sum_{\sigma\in \mathcal{P}_{fi}(t)}D_{\rm p}^{\pi}\left[s_{\sigma}(t)\right]+
	\sum_{i\in\mathcal{V}}\lambda_{i}C+\sum_{i\in\mathcal{V}}\mu_iP,
\end{align}
where $\lambda_i$ and $\mu_i$ are the Lagrangian multipliers.
Ignore the constant value, it can be further derived as
\begin{align}\notag
	&\mathfrak{L}(\pi,\lambda,\mu) =\lim_{T\to \infty} \frac{1}{T} \sum_{t=0}^{T-1}\sum_{f\in\mathcal{F}}\left\{\sum_{\sigma\in \mathcal{P}_f(t)}w_fR_{r}^{\pi}\left[s_{\sigma}(t)\right]-\right.\\ \notag
	&\left.\sum_{i\in\mathcal{V}}\lambda_i\sum_{\sigma\in \mathcal{P}_{fi}(t)\cup \mathcal{P}_{\mathcal{I}_i}(t)}D_{\rm c}^{\pi}\left[s_{\sigma}(t)\right]-\sum_{i\in\mathcal{V}}\mu_i\sum_{\sigma\in \mathcal{P}_{fi}(t)}D_{\rm p}^{\pi}\left[s_{\sigma}(t)\right]\right\}\\ \notag
	&=\lim_{T\to \infty} \frac{1}{T} \sum_{t=0}^{T-1}\sum_{f\in\mathcal{F}}\sum_{i\in\mathcal{V}}\left\{\sum_{\sigma\in \mathcal{P}_{fi}(t)}w_fR_{r}^{\pi}\left[s_{\sigma}(t)\right]-\right.\\ \notag
	&\left.\lambda_i\sum_{\sigma\in \mathcal{P}_{fi}(t)\cup \mathcal{P}_{\mathcal{I}_i}(t)}D_{\rm c}^{\pi}\left[s_{\sigma}(t)\right]-\mu_i\sum_{\sigma\in \mathcal{P}_{fi}(t)}D_{\rm p}^{\pi}\left[s_{\sigma}(t)\right]\right\}\\ \notag
	&=\lim_{T\to \infty} \frac{1}{T} \sum_{t=0}^{T-1}\sum_{f\in\mathcal{F}}\sum_{i\in\mathcal{V}}\left\{\sum_{\sigma\in \mathcal{P}_{fi}(t)}\left[w_fR_{r}^{\pi}\left[s_{\sigma}(t)\right]-\right.\right.\\ \notag
	&\left.\left.\lambda_iD_{\rm c}^{\pi}\left[s_{\sigma}(t)\right]-\mu_iD_{\rm p}^{\pi}\left[s_{\sigma}(t)\right]\right]-\lambda_i\sum_{\sigma\in  \mathcal{P}_{\mathcal{I}_i}(t)}D_{\rm c}^{\pi}\left[s_{\sigma}(t)\right]\right\}.
\end{align}
Note that the last term is not related to $\sigma$ in $\mathcal{P}_{fi}(t)$.
Define the number of packet at node $i$ is $N_i$, then we can get
\begin{align}
	\sum_{\sigma\in  \mathcal{P}_{\mathcal{I}_i}(t)}D_{\rm c}^{\pi}\left[s_{\sigma}(t)\right]=\frac{1}{N_i}\sum_{\sigma\in \mathcal{P}_{fi}(t)}\sum_{\sigma\in  \mathcal{P}_{\mathcal{I}_i}(t)}D_{\rm c}^{\pi}\left[s_{\sigma}(t)\right].
\end{align}
Therefore, the Lagrangian is derived as
\begin{align}\notag
	&\mathfrak{L}(\pi,\lambda,\mu) =\lim_{T\to \infty} \frac{1}{T} \sum_{t=0}^{T-1}\sum_{f\in\mathcal{F}}\sum_{i\in\mathcal{V}}\sum_{\sigma\in \mathcal{P}_{fi}(t)}\left[w_fR_{r}^{\pi}\left[s_{\sigma}(t)\right]-\right.\\ \notag
	&\left.\lambda_iD_{\rm c}^{\pi}\left[s_{\sigma}(t)\right]-\mu_iD_{\rm p}^{\pi}\left[s_{\sigma}(t)\right]-\frac{\lambda_i}{N_i}\sum_{\sigma\in  \mathcal{P}_{\mathcal{I}_i}(t)}D_{\rm c}^{\pi}\left[s_{\sigma}(t)\right]\right]\\ \notag
	&=\lim_{T\to \infty} \frac{1}{T} \sum_{t=0}^{T-1}\sum_{f\in\mathcal{F}}\sum_{i\in\mathcal{V}}\sum_{\sigma\in \mathcal{P}_{fi}(t)}R_{r_{\rm L}}^{\pi,\lambda,\mu}\left[s_{\sigma}(t)\right],
\end{align}
where $R_{r_{\rm L}}^{\pi,\lambda,\mu}\left[s_{\sigma}(t)\right]$ is to replace $r(\cdot)$ in \eqref{eq:ECReward} with $r_{\rm L}(\cdot)$.
$r_{\rm L}(\cdot)$ is the reward under the Lagrange multipliers and is given by
\begin{align}
	r_{\rm L}^{\sigma}(t) = w_fr^{\sigma}(t)-\lambda_ic^{\sigma}(t)-\mu_{i}p^{\sigma}(t)-\frac{\lambda_i}{N_i}\sum_{\sigma^{\prime}\in\mathcal{P}_{\mathcal{I}_i}(t)}c^{\sigma^{\prime}}(t),
	\label{eq:RewardL}
\end{align}
where $r^{\sigma}(t)$, $c^{\sigma}(t)$, and $p^{\sigma}(t)$ are simplified representations of $r(\cdot)$ in \eqref{eq:ECReward}, $c(\cdot)$ in \eqref{eq:subcarrierCons}, and $p(\cdot)$ in \eqref{eq:powerCons}.

\section{Proof of Theorem 1} \label{apdx:AuxProof}
Without considering the auxiliary reward, the cumulative reward of any state $s_{\sigma}(t)$ is $R_{r_{\rm D}}^{\pi}\left[s_{\sigma}(t)\right]$ and
\begin{align}
	R_{r_{\rm D}}^{\pi}\left[s_{\sigma}(t)\right]=\sum_{a_{\sigma}(t)\in\mathcal{A}}\pi[a_{\sigma}(t)|s_{\sigma}(t)]D_{r_{\rm D}}^{\pi}[s_{\sigma}(t),a_{\sigma}(t)].
\end{align}
Under the optimal policy, $D_{r_{\rm D}}^{*}[s_{\sigma}(t),a_{\sigma}(t)]$ should satisfy the Bellman equation, that is
\begin{align}\notag
	D_{r_{\rm D}}^{*}[s_{\sigma}(t),a_{\sigma}(t)]=&\mathbb{E}_{\pi}\left\{r_{\rm D}[s_{\sigma}(t),a_{\sigma}(t),s_{\sigma}(t+1)]+\right.\\ \notag
	&\left.\max_{a_{\sigma}(t+1)\in\mathcal{A}}D_{r_{\rm D}}^{*}[s_{\sigma}(t+1),a_{\sigma}(t+1)]\right\}.
\end{align}
Suppose the potential function is $\frac{u_{\rm L}\left[s_{\sigma}(t)\right]}{d_{\sigma i}}$.
For the convenience of expression, let $s = s_{\sigma}(t)=(i,\tau)$ and $s^{\prime} = s_{\sigma}(t+1)=(j,\tau-1)$.
It can be derived that
\begin{align}\notag
	&D_{r_{\rm D}}^{*}[s,a]-\frac{u_{\rm L}\left[s\right]}{d_{\sigma i}} = \mathbb{E}_{\pi}\left\{r_{\rm D}[s,a,s^{\prime}]+\max_{a^{\prime}\in\mathcal{A}}D_{r_{\rm D}}^{*}[s^{\prime},a^{\prime}]\right\}-\frac{u_{\rm L}\left[s\right]}{d_{\sigma i}}\\ \notag
	&=\mathbb{E}_{\pi}\left\{r_{\rm D}[s,a,s^{\prime}]+\frac{u_{\rm L}\left[s^{\prime}\right]}{d_{\sigma j}}+\max_{a^{\prime}\in\mathcal{A}}D_{r_{\rm D}}^{*}[s^{\prime},a^{\prime}]-\frac{u_{\rm L}\left[s^{\prime}\right]}{d_{\sigma j}}\right\}-\frac{u_{\rm L}\left[s\right]}{d_{\sigma i}}\\ \notag
	&=\mathbb{E}_{\pi}\left\{r_{\rm D}[s,a,s^{\prime}]+\frac{u_{\rm L}\left[s^{\prime}\right]}{d_{\sigma j}}-\frac{u_{\rm L}\left[s\right]}{d_{\sigma i}}+\max_{a^{\prime}\in\mathcal{A}}D_{r_{\rm D}}^{*}[s^{\prime},a^{\prime}]-\frac{u_{\rm L}\left[s^{\prime}\right]}{d_{\sigma j}}\right\}\\ \notag
	&=\mathbb{E}_{\pi}\left\{r[s,a,s^{\prime}]+\max_{a^{\prime}\in\mathcal{A}}D_{r_{\rm D}}^{*}[s^{\prime},a^{\prime}]-\frac{u_{\rm L}\left[s^{\prime}\right]}{d_{\sigma j}}\right\}.
\end{align}
Let $D_{r}^{\pi}[s,a]=D_{r_{\rm D}}^{*}[s,a]-\frac{u_{\rm L}\left[s\right]}{d_{\sigma i}}$, then we can get
\begin{align}
	D_{r}^{\pi}[s,a]=\mathbb{E}_{\pi}\left\{r[s,a,s^{\prime}]+\max_{a^{\prime}\in\mathcal{A}}D_{r}^{\pi}[s^{\prime},a^{\prime}]\right\}.
\end{align}
Therefore, for any state, $D_{r}^{\pi}[s,a]$ still satisfies the Bellman equation then it is also optimal.
Finally, the optimal policy is obtained as that without auxiliary reward.

\section{Policy gradient derivation} \label{apdx:PolicyGradientDerive}
Recall that the Lagrangian is given by
\begin{align}
	\mathfrak{L}\left ( \pi,\lambda,\mu \right ) = \lim_{T\to \infty} \frac{1}{T} \sum_{t=0}^{T-1}\sum_{f\in\mathcal{F}}\sum_{i\in\mathcal{V}}\sum_{\sigma\in \mathcal{P}_{fi}(t)}R_{r_{\rm L}}^{\pi,\lambda,\mu}\left[s_{\sigma}(t)\right]. 
\end{align}
A parameterized policy $\pi_{\theta}$ is used to obtained the gradient, such as $\pi_{\theta}(a|s)=\frac{\exp(\theta_{sa})}{\sum_{a^{\prime}\in \mathcal{A}}\exp(\theta_{s,a^{\prime}})}$, which is differentiable and tractable.
Utilizing the policy gradient theorem, the gradient for the objective function is
\begin{align}\notag
	&\nabla_{\theta}\mathfrak{L}\left ( \pi,\lambda,\mu \right ) \\ \notag
	&= \lim_{T\to \infty} \frac{1}{T} \sum_{t=0}^{T-1}\sum_{f\in\mathcal{F}}\sum_{i\in\mathcal{V}}\sum_{\sigma\in \mathcal{P}_{fi}(t)}&\nabla_{\theta}\log \pi_{\theta}[a_{\sigma}(t)|s_{\sigma}(t)]\\ \notag
	& &D^{\pi,\lambda,\mu}_{r_{\rm L}}[s_{\sigma}(t),a_{\sigma}(t)],
\end{align}
where
\begin{align}\notag
	&D^{\pi,\lambda,\mu}_{r_{\rm L}}[s_{\sigma}(t),a_{\sigma}(t)]\\
	&=\mathbb{E}_{\pi}\left\{\left.\sum_{k=0}^{\tau_f-1}r_{\rm L}^{\pi,\lambda,\mu}\left[s_{\sigma}(t+k),a_{\sigma}(t+k),s_{\sigma}(t+k+1)\right]\right|s_{\sigma}(t),a_{\sigma}(t)\right\}.
\end{align}
Then, a general policy gradient updating method is $\theta_{sa}(t+1) = \theta_{sa}(t)+\eta_1\nabla_{\theta}\mathfrak{L}\left ( \pi,\lambda,\mu \right )|_{s,a}$.
Nowadays, the natural policy gradient (NPG) method is popular and has a faster convergence rate than other policy gradient methods, in which the policy is updated as
\begin{align}
	\theta_{sa}(t+1) = \theta_{sa}(t)+\eta_1\mathcal{F}^{\dagger}\left[\theta_{sa}(t)\right]\left.\nabla_{\theta}\mathfrak{L}\left ( \pi,\lambda,\mu \right )\right|_{s,a},
\end{align}
where $\mathcal{F}(\theta)$ is the Fisher information matrix and given by
\begin{align}
	\mathcal{F}(\theta) = \mathbb{E}_{\pi}\left[\nabla_{\theta}\log \pi_{\theta}(a|s)\left(\nabla_{\theta}\log \pi_{\theta}(a|s)\right)^{\top}\right].
\end{align}
$\mathcal{F}^{\dagger}$ takes the Moore-Penrose inverse of matrix $\mathcal{F}$.
Using the compatible function approximation error theorem, the policy gradient is derived as
\begin{align}\notag
	 &\mathcal{F}^{\dagger}\left[\theta_{sa}(t)\right]\left.\nabla_{\theta}\mathfrak{L}\left ( \pi,\lambda,\mu \right )\right|_{s,a} \\
	 &= D^{\pi,\lambda,\mu}_{r_{\rm L}}[s_{\sigma}(t)=s,a_{\sigma}(t)=a] - R_{r_{\rm L}}^{\pi,\lambda,\mu}\left[s_{\sigma}(t)=s\right].
\end{align}
By using the exponential relationship between policy $\pi$ and $\theta$, the update of the policy can be obtained as
\begin{align}
	\pi_{\theta_{sa}}(t)=\frac{\pi_{\theta_{sa}}(t-1)e^{\eta_1 \left.\nabla_{\theta}\mathfrak{L}\left ( \pi,\lambda,\mu \right )\right|_{s_{\sigma}(t-1),a_{\sigma}(t-1)}}}{\sum_{a\in \mathcal{A}_i}\pi_{\theta_{sa}}(t)e^{ \eta_1 \left.\nabla_{\theta}\mathfrak{L}\left ( \pi,\lambda,\mu \right )\right|_{s_{\sigma}(t-1),a}}}.
\end{align}

\section{Proof of Theorem 2} \label{apdx:ProofTheorem2}
\begin{align}\notag
	\frac{1}{T} \sum_{t=0}^{T-1}\sum_{f\in\mathcal{F}}\sum_{\sigma\in \mathcal{P}_f(t)}  w_f \left\{ R^{*}_{r}\left[s_{\sigma}(t)\right]-R^{\pi}_{r}\left[s_{\sigma}(t)\right] \right\}\leq 
	\frac{g(N,F,C,\tau_{\max})}{T},
\end{align}

Recall the objective function is 
\begin{align}\notag
	\frac{1}{T} \sum_{t=0}^{T-1}\sum_{f\in\mathcal{F}}\sum_{\sigma\in \mathcal{P}_f(t)}  w_f  R^{\pi}_{r}\left[s_{\sigma}(t)\right].
\end{align}
The optimal reward is $R^{*}_{r}\left[s_{\sigma}(t)\right]$.
The optimality gap is 
\begin{align}\notag
	&\frac{1}{T} \sum_{t=0}^{T-1}\sum_{f\in\mathcal{F}}\sum_{\sigma\in \mathcal{P}_f(t)}  w_f \left\{ R^{*}_{r}\left[s_{\sigma}(t)\right]-R^{\pi}_{r}\left[s_{\sigma}(t)\right] \right\}\\
	&=\frac{1}{T} \sum_{t=0}^{T-1} \mathbb{E}_{\pi^*}\left[\sum_{a_{\sigma}(t)\in \mathcal{A}} \pi^{*}[a_{\sigma}(t)|s_{\sigma}(t)]A^{\pi}_r[s_{\sigma}(t),a_{\sigma}(t)]\right],
	\label{eq:performancediff}
\end{align}
where $A^{\pi}_r[s_{\sigma}(t),a_{\sigma}(t)]$ is derived by the performance difference theorem and is denoted by
\begin{align}
	A^{\pi}_r[s_{\sigma}(t),a_{\sigma}(t)]=D^{\pi}_{r}[s_{\sigma}(t),a_{\sigma}(t)] - R_{r}^{\pi}\left[s_{\sigma}(t)\right].
\end{align}
According to the PGDS method in Algorithm \ref{alg:algorithm1}, it should be noted that 
\begin{align}
	\pi_{\theta_{sa}}(t+1) = \pi_{\theta_{sa}}(t)\exp \left\{\eta_1 A_{r_{\rm L}}^{\pi,\lambda,\mu}[s_{\sigma}(t),a_{\sigma}(t)] \right\}Z^{-1}[s_{\sigma}(t)],\\ 	\label{eq:policyparagradient} \notag
	A_{r_{\rm L}}^{\pi,\lambda,\mu}[s_{\sigma}(t),a_{\sigma}(t)] = A^{\pi}_r[s_{\sigma}(t),a_{\sigma}(t)] + \lambda_{i}(t) A^{\pi}_c[s_{\sigma}(t),a_{\sigma}(t)] + \\
	\mu_{i}(t) A^{\pi}_p[s_{\sigma}(t),a_{\sigma}(t)],
\end{align}
where $Z$ is the normalized function, $A^{\pi}_c[s_{\sigma}(t),a_{\sigma}(t)]$ is the function related to the subcarriers and $A^{\pi}_p[s_{\sigma}(t),a_{\sigma}(t)]$ is the function related to the power constraint, which are defined in the same way as $A^{\pi}_r[s_{\sigma}(t),a_{\sigma}(t)]$.
Then, it is derived that
\begin{align}\notag
	A_{r}^{\pi,\lambda,\mu}[s_{\sigma}(t),a_{\sigma}(t)] = \frac{1}{\eta_1} \log \frac{\pi_{\theta_{sa}}(t+1)Z[s_{\sigma}(t)]}{\pi_{\theta_{sa}}(t)}-\\
	\lambda_{i}(t) A^{\pi}_c[s_{\sigma}(t),a_{\sigma}(t)]-\mu_{i}(t) A^{\pi}_p[s_{\sigma}(t),a_{\sigma}(t)].
\end{align}
Following the derivation in \eqref{eq:performancediff}, it is obtained that
\begin{align} \notag
	&\mathbb{E}_{\pi^*}\left[\sum_{a_{\sigma}(t)\in \mathcal{A}} \pi^{*}[a_{\sigma}(t)|s_{\sigma}(t)]A^{\pi}_r[s_{\sigma}(t),a_{\sigma}(t)]\right]\\ \notag
	&=\frac{1}{\eta_1}\mathbb{E}_{\pi^*}\left[\sum_{a_{\sigma}(t)\in \mathcal{A}} \pi^{*}[a_{\sigma}(t)|s_{\sigma}(t)]\log \frac{\pi_{\theta_{sa}}(t+1)Z[s_{\sigma}(t)]}{\pi_{\theta_{sa}}(t)}\right]-\\ \notag
	&\mathbb{E}_{\pi^*}\left[\sum_{a_{\sigma}(t)\in \mathcal{A}}\lambda_i(t) \pi^{*}[a_{\sigma}(t)|s_{\sigma}(t)]A^{\pi}_c[s_{\sigma}(t),a_{\sigma}(t)]\right]-\\
	&\mathbb{E}_{\pi^*}\left[\sum_{a_{\sigma}(t)\in \mathcal{A}}\mu_i(t) \pi^{*}[a_{\sigma}(t)|s_{\sigma}(t)]A^{\pi}_p[s_{\sigma}(t),a_{\sigma}(t)]\right]
\end{align}
Using the performance difference theorem for two constraints, it can be obtained that
\begin{align}\notag
	\mathbb{E}_{\pi^*}\left[\sum_{a_{\sigma}(t)\in \mathcal{A}}\lambda_i(t) \pi^{*}[a_{\sigma}(t)|s_{\sigma}(t)]A^{\pi}_c[s_{\sigma}(t),a_{\sigma}(t)]\right] \\ \notag
	= \mathbb{E}_{\pi^*}\left[\lambda_i(t)c[s_{\sigma}(t),a_{\sigma}(t)]\right]-\mathbb{E}_{\pi}\left[\lambda_i(t)c[s_{\sigma}(t),a_{\sigma}(t)]\right],\\ \notag
	\mathbb{E}_{\pi^*}\left[\sum_{a_{\sigma}(t)\in \mathcal{A}}\mu_i(t) \pi^{*}[a_{\sigma}(t)|s_{\sigma}(t)]A^{\pi}_p[s_{\sigma}(t),a_{\sigma}(t)]\right] \\ \notag
	= \mathbb{E}_{\pi^*}\left[\mu_i(t)p[s_{\sigma}(t),a_{\sigma}(t)]\right]-\mathbb{E}_{\pi}\left[\mu_i(t)p[s_{\sigma}(t),a_{\sigma}(t)]\right].
\end{align}
Then \eqref{eq:performancediff} is derived as
\begin{align}\notag
	&\frac{1}{T} \sum_{t=0}^{T-1} \mathbb{E}_{\pi^*}\left[\sum_{a_{\sigma}(t)\in \mathcal{A}} \pi^{*}[a_{\sigma}(t)|s_{\sigma}(t)]A^{\pi}_r[s_{\sigma}(t),a_{\sigma}(t)]\right]\\ \notag
	&=\frac{1}{\eta_1T}\sum_{t=0}^{T-1}\mathbb{E}_{\pi^*}\left[\sum_{a_{\sigma}(t)\in \mathcal{A}} \pi^{*}[a_{\sigma}(t)|s_{\sigma}(t)]\log \frac{\pi_{\theta_{sa}}(t+1)Z[s_{\sigma}(t)]}{\pi_{\theta_{sa}}(t)}\right]-\\ \notag
	&\sum_{t=0}^{T-1}\mathbb{E}_{\pi^*}\left[\lambda_i(t)c[s_{\sigma}(t),a_{\sigma}(t)]\right]-\mathbb{E}_{\pi}\left[\lambda_i(t)c[s_{\sigma}(t),a_{\sigma}(t)]\right]-\\ 
	&\sum_{t=0}^{T-1}\mathbb{E}_{\pi^*}\left[\mu_i(t)p[s_{\sigma}(t),a_{\sigma}(t)]\right]-\mathbb{E}_{\pi}\left[\mu_i(t)p[s_{\sigma}(t),a_{\sigma}(t)]\right].
	\label{eq:derivations}
\end{align}
The first term is derived as 
\begin{align}\notag
	\label{eq:firstterm}
	&\frac{1}{\eta_1T}\sum_{t=0}^{T-1}\mathbb{E}_{\pi^*}\left[\sum_{a_{\sigma}(t)\in \mathcal{A}} \pi^{*}[a_{\sigma}(t)|s_{\sigma}(t)]\log \frac{\pi_{\theta_{sa}}(t+1)Z[s_{\sigma}(t)]}{\pi_{\theta_{sa}}(t)}\right]\\ \notag
	&=\frac{1}{\eta_1T}\sum_{t=0}^{T-1}\mathbb{E}_{\pi^*}\left[\sum_{a_{\sigma}(t)\in \mathcal{A}} \pi^{*}[a_{\sigma}(t)|s_{\sigma}(t)]\log \frac{\pi_{\theta_{sa}}(t+1)}{\pi_{\theta_{sa}}(t)}\right]+\\ \notag
	&\frac{1}{\eta_1T}\sum_{t=0}^{T-1}\mathbb{E}_{\pi^*}\left[\sum_{a_{\sigma}(t)\in \mathcal{A}} \pi^{*}[a_{\sigma}(t)|s_{\sigma}(t)]\log Z[s_{\sigma}(t)]\right]\\ \notag
	&=\frac{1}{\eta_1T}\mathbb{E}_{\pi^*}\left[\log \frac{\pi_{\theta_{sa}}(T)}{\pi_{\theta_{sa}}(0)}\right]+\frac{1}{\eta_1T}\sum_{t=0}^{T-1}\mathbb{E}_{\pi^*}\left[\log Z[s_{\sigma}(t)]\right]\\ 
	&\leq \frac{\log NFC\tau_{\max}}{\eta_1T} \\ \notag
	&+\frac{1}{T}\sum_{t=0}^{T-1}\left\{\mathbb{E}_{\pi^{*}}\left[r[s_{\sigma}(t+1),a_{\sigma}(t+1)]\right]-\mathbb{E}_{\pi^*}\left[r[s_{\sigma}(t),a_{\sigma}(t)]\right]\right\} \\ \notag
	&+ \frac{1}{T}\sum_{t=0}^{T-1} \left\{\mathbb{E}_{\pi^{*}}\left[\lambda_{i}(t+1)c[s_{\sigma}(t+1),a_{\sigma}(t+1)]\right]-\right.\\ \notag
	&\left.\mathbb{E}_{\pi^*}\left[\lambda_{i}(t)c[s_{\sigma}(t),a_{\sigma}(t)]\right]\right\}\\ \notag
	&+\frac{1}{T}\sum_{t=0}^{T-1} \left\{\mathbb{E}_{\pi^{*}}\left[\mu_{i}(t+1)p[s_{\sigma}(t+1),a_{\sigma}(t+1)]\right]-\right.\\ \notag
	&\left.\mathbb{E}_{\pi^*}\left[\mu_{i}(t)p[s_{\sigma}(t),a_{\sigma}(t)]\right]\right\} \\ \notag 
	&\leq \frac{\log NFC\tau_{\max}}{\eta_1T} + \frac{\bar{r}_{T}}{T} +\frac{\bar{\lambda}\bar{c}+\bar{\mu}\bar{p}}{T} -\frac{\mathbb{E}_{\pi}\left(\lambda_i^2(T)\right)-\epsilon_{\lambda} }{2\eta_2T}-\frac{\mathbb{E}_{\pi}\left(\mu_i^2(T)\right)-\epsilon_{\mu} }{2\eta_3T}.
\end{align}
where the fact that $\mathbb{E}(\log \frac{p(\pi)}{q(\pi)})\leq \log |\pi|$ is used and $|\pi|$ means the size of $\pi$.
The maximal size of $\pi$ for flow $f$ at node $i$ is $C\tau_{\max}$, then the total policy for the whole network is $NFC\tau_{\max}$.
The inequality in \eqref{eq:firstterm} uses the expansion of $ \mathbb{E}_{\pi^*}\left[\log Z[s_{\sigma}(t)]\right]$, which is derived as
\begin{align}\notag
	&\frac{1}{\eta_1}\mathbb{E}_{\pi^*}\left[\log Z[s_{\sigma}(t)]\right]\leq\frac{1}{\eta_1}\mathbb{E}_{\pi^*}\left[\log Z[s_{\sigma}(t)]\right]+\\ \notag
	&\frac{1}{\eta_1}\mathbb{E}_{\pi^*}\left[\sum_{a_{\sigma}(t)\in \mathcal{A}} \pi^{*}[a_{\sigma}(t)|s_{\sigma}(t)]\log \frac{\pi_{\theta_{sa}}(t+1)}{\pi_{\theta_{sa}}(t)}\right]\\ \notag
	&=\frac{1}{\eta_1}\mathbb{E}_{\pi^*}\left[\sum_{a_{\sigma}(t)\in \mathcal{A}} \pi^{*}[a_{\sigma}(t)|s_{\sigma}(t)]\log \frac{\pi_{\theta_{sa}}(t+1)Z[s_{\sigma}(t)]}{\pi_{\theta_{sa}}(t)}\right]-\\ \notag
	&\mathbb{E}_{\pi^*}\left[\lambda_{i}(t)A^{\pi}_c[s_{\sigma}(t),a_{\sigma}(t)]\right] - \mathbb{E}_{\pi^*}\left[\mu_{i}(t)A^{\pi}_p[s_{\sigma}(t),a_{\sigma}(t)]\right]+\\ \notag
	&\mathbb{E}_{\pi^*}\left[\lambda_{i}(t)A^{\pi}_c[s_{\sigma}(t),a_{\sigma}(t)]\right] + \mathbb{E}_{\pi}\left[\mu_{i}(t)A^{\pi}_p[s_{\sigma}(t),a_{\sigma}(t)]\right]\\ \notag
	&=\mathbb{E}_{\pi^*}\left[A^{\pi}_{r}[s_{\sigma}(t),a_{\sigma}(t)]\right]+\mathbb{E}_{\pi^*}\left[\lambda_{i}(t)A^{\pi}_c[s_{\sigma}(t),a_{\sigma}(t)]\right] +\\ \notag
	&\mathbb{E}_{\pi^*}\left[\mu_{i}(t)A^{\pi}_p[s_{\sigma}(t),a_{\sigma}(t)]\right]\\ \notag
	&=\mathbb{E}_{\pi^{*}}\left[r[s_{\sigma}(t+1),a_{\sigma}(t+1)]\right]-\mathbb{E}_{\pi^*}\left[r[s_{\sigma}(t),a_{\sigma}(t)]\right] \\ \notag
	&+  \left\{\mathbb{E}_{\pi^{*}}\left[\lambda_{i}(t+1)c[s_{\sigma}(t+1),a_{\sigma}(t+1)]\right]-\mathbb{E}_{\pi^*}\left[\lambda_{i}(t)c[s_{\sigma}(t),a_{\sigma}(t)]\right]\right\}\\ \notag
	&+ \left\{\mathbb{E}_{\pi^{*}}\left[\mu_{i}(t+1)p[s_{\sigma}(t+1),a_{\sigma}(t+1)]\right]-\mathbb{E}_{\pi^*}\left[\mu_{i}(t)p[s_{\sigma}(t),a_{\sigma}(t)]\right]\right\}
\end{align}
The third term of the inequality in \eqref{eq:firstterm} is derived by
\begin{align}
	&\frac{1}{T}\sum_{t=0}^{T-1} \left\{\mathbb{E}_{\pi^{*}}\left[\lambda_{i}(t+1)c[s_{\sigma}(t+1),a_{\sigma}(t+1)]\right]-\mathbb{E}_{\pi^*}\left[\lambda_{i}(t)c[s_{\sigma}(t),a_{\sigma}(t)]\right]\right\}\\ \notag
	&=\frac{1}{T}\sum_{t=0}^{T-1} \mathbb{E}_{\pi^{*}}\left\{\lambda_{i}(t+1)c[s_{\sigma}(t+1),a_{\sigma}(t+1)]-\lambda_{i}(t)c[s_{\sigma}(t),a_{\sigma}(t)]\right\}\\ \notag
	&=\frac{1}{T}\mathbb{E}_{\pi^{*}}\left\{\sum_{t=0}^{T-1}\lambda_{i}(t+1)c[s_{\sigma}(t+1),a_{\sigma}(t+1)]-\lambda_{i}(t)c[s_{\sigma}(t),a_{\sigma}(t)]\right\}\\ \notag
	&=\frac{1}{T}\mathbb{E}_{\pi^{*}}\left\{\lambda_{i}(T)c[s_{\sigma}(T),a_{\sigma}(T)]-\lambda_{i}(0)c[s_{\sigma}(0),a_{\sigma}(0)]\right\}\\ \notag
	&=\frac{\mathbb{E}_{\pi^{*}}\left\{\lambda_{i}(T)c[s_{\sigma}(T),a_{\sigma}(T)]\right\}}{T}.
\end{align}
The fourth term of the inequality in \eqref{eq:firstterm} is similar to the third term. 
Following the same derivation process, it can be derived that 
\begin{align}
	\frac{1}{T}\sum_{t=0}^{T-1}\notag \left\{\mathbb{E}_{\pi^{*}}\left[\mu_{i}(t+1)p[s_{\sigma}(t+1),a_{\sigma}(t+1)]\right]-\mathbb{E}_{\pi^*}\left[\mu_{i}(t)p[s_{\sigma}(t),a_{\sigma}(t)]\right]\right\} \\
	=\frac{\mathbb{E}_{\pi^{*}}\left\{\mu_{i}(T)p[s_{\sigma}(T),a_{\sigma}(T)]\right\}}{T}.
\end{align}

Before deriving the second term of \eqref{eq:derivations}, it should be noted that the update of Lagrangian multiplier $\lambda$ is
\begin{align}
	\lambda_{i}(t+1)=\left(\lambda_{i}(t)+\eta_2\mathbb{E}_{\pi}\left[C-c[s_{\sigma}(t),a_{\sigma}(t)]\right]\right)_{+},
\end{align}
where $()_{+}$ denotes the projection into an interval $\Lambda_1$ on the positive real set $\mathbb{R}_{+}$ and $\eta_2$ is the step size.
Since the projection makes the range of $\lambda$ smaller, it is obtained that $\lambda_{i}(t+1)-\lambda_{i}(t)\leq \eta_2 \mathbb{E}_{\pi^*}\left[C-c[s_{\sigma}(t),a_{\sigma}(t)]\right]$.
Note that $\mathbb{E}_{\pi^*}\left[c[s_{\sigma}(t),a_{\sigma}(t)]\right] = C$, the second term of \eqref{eq:derivations} is derived as 
\begin{align}\notag
	&\frac{1}{T} \sum_{t=0}^{T-1}\left\{\mathbb{E}_{\pi^*}\left[\lambda_{i}(t)c[s_{\sigma}(t),a_{\sigma}(t)]\right]-\mathbb{E}_{\pi}\left[\lambda_{i}(t)c[s_{\sigma}(t),a_{\sigma}(t)]\right]\right\}\\ \notag
	&\geq \frac{1}{\eta_2T} \sum_{t=0}^{T-1} \mathbb{E}_{\pi}\left\{\lambda_{i}(t) \left[\lambda_{i}(t+1)-\lambda_{i}(t)\right]\right\}\\ \notag
	&=\frac{1}{\eta_2T} \left[\frac{1}{2}\sum_{t=0}^{T-1}\mathbb{E}_{\pi}\left(\lambda_{i}^2(t+1)-\lambda_{i}^2(t)\right)-\frac{1}{2}\sum_{t=0}^{T-1}\mathbb{E}_{\pi}\left(\lambda_{i}(t+1)-\lambda_{i}(t)\right)^2\right]\\ \notag
	&=\frac{1}{\eta_2T} \left[\frac{1}{2}\mathbb{E}_{\pi}\left(\lambda_i^2(T)\right)-\frac{1}{2}\sum_{t=0}^{T-1}\mathbb{E}_{\pi}\left(\lambda_{i}(t+1)-\lambda_{i}(t)\right)^2\right]\\
	&=\frac{\mathbb{E}_{\pi}\left(\lambda_i^2(T)\right)-\epsilon_{\lambda} }{2\eta_2T},
\end{align}
where $\epsilon$ is the quadratic difference of the iterative step.
As the iterations go on, $\lambda_{i}(t+1)$ and $\lambda_{i}(t)$ tend to be equal, so $\epsilon_{\lambda}$ is a small enough finite value.

The third term of \eqref{eq:derivations} has a similar format as the second term, so it can be derived that
\begin{align}\notag
	\frac{1}{T} \sum_{t=0}^{T-1}\left\{\mathbb{E}_{\pi^*}\left[\mu_{i}(t)p[s_{\sigma}(t),a_{\sigma}(t)]\right]-\mathbb{E}_{\pi}\left[\mu_{i}(t)p[s_{\sigma}(t),a_{\sigma}(t)]\right]\right\}\\ 
	\geq \frac{\mathbb{E}_{\pi}\left(\mu_i^2(T)\right)-\epsilon_{\mu} }{2\eta_3T}.
\end{align}
It should be noted that $\epsilon_{\mu}$ is similar as $\epsilon_c$.
Finally, it can be obtained 
\begin{align}\notag
	& \frac{1}{T} \sum_{t=0}^{T-1}\sum_{f\in\mathcal{F}}\sum_{\sigma\in \mathcal{P}_f(t)}  w_f \left\{ R^{\pi}_{r}\left[s_{\sigma}(t)\right]-R^{*}_{r\sigma} \right\}\leq \frac{\log NFC\tau_{\max}}{\eta_1T} + \frac{\bar{r}_{T}}{T} +\\ \notag &\frac{\mathbb{E}_{\pi^{*}}\left\{\lambda_{i}(T)c[s_{\sigma}(T),a_{\sigma}(T)]+\mu_{i}(T)p[s_{\sigma}(T),a_{\sigma}(T)]\right\}}{T} - \\
	&\frac{\mathbb{E}_{\pi}\left(\lambda_i^2(T)\right)-\epsilon_{\lambda} }{2\eta_2T}-\frac{\mathbb{E}_{\pi}\left(\mu_i^2(T)\right)-\epsilon_{\mu} }{2\eta_3T}.
\end{align}
For simplicity, the third term of the above expressions on the right-hand side is denoted as $\frac{\bar{\lambda}\bar{c}+\bar{\mu}\bar{p}}{T}$.
To clearly reflect the influence of network parameters on optimality, the above formula can be expressed as
\begin{align}
	\frac{1}{T} \sum_{t=0}^{T-1}\sum_{f\in\mathcal{F}}\sum_{\sigma\in \mathcal{P}_f(t)}  w_f \left\{ R^{\pi}_{r}\left[s_{\sigma}(t)\right ]-R^{*}_{r\sigma} \right\}  \leq \frac{g(N,F,C,\tau_{\max})}{T}
\end{align}
Therefore, as $T$ approaches infinity, the optimality gap goes to zero, that is 
\begin{align}
	\frac{1}{T} \sum_{t=0}^{T-1}\sum_{f\in\mathcal{F}}\sum_{\sigma\in \mathcal{P}_f(t)}  w_f \left\{ R^{\pi}_{r}\left[s_{\sigma}(t)\right ]-R^{*}_{r\sigma} \right\}= \mathcal{O}\left(\frac{1}{T}\right).
\end{align}

\section{Proof of Theorem 3} \label{apdx:ProofTheorem3}
\begin{align}\notag
	\frac{1}{T} \sum_{t=0}^{T-1}\sum_{f\in\mathcal{F}}\sum_{\sigma\in \mathcal{P}_f(t)}  w_f \left\{ R^{\pi}_{r}\left[s_{\sigma}(t)\right ]-R^{*}_{r} \right\}\leq \\
	\frac{g(N,F,C,\tau_{\max})+q(\delta_c,\delta_p)}{T}.
\end{align}

Recall the number and power of subcarriers with capacity unavailable condition is $\tilde{c}^{\sigma}(t)$ and $\tilde{p}^{\sigma}(t)$, the policy is denoted as $\tilde{\pi}$ and the parameterized policy is denoted as $\tilde{\theta}$.
The optimality gap between the optimal policy $\pi_{*}$ and the policy $\pi$ with capacity unavailable condition is
\begin{align}\notag
	&\frac{1}{T} \sum_{t=0}^{T-1}\sum_{f\in\mathcal{F}}\sum_{\sigma\in \mathcal{P}_f(t)}  w_f \left\{ R^{\pi}_{\tilde{r}}\left[s_{\sigma}(t)\right]-R^{*}_{r}\left[s_{\sigma}(t)\right] \right\}\\
	&=\frac{1}{T} \sum_{t=0}^{T-1} \mathbb{E}_{\pi^*}\left[\sum_{a_{\sigma}(t)\in \mathcal{A}} \pi^{*}[a_{\sigma}(t)|s_{\sigma}(t)]A^{\pi}_{\tilde{r}}[s_{\sigma}(t),a_{\sigma}(t)]\right],
\end{align}
For the convenience of derivation, using $s=s_{\sigma}(t)=(i,\tau)$ and $a=a_{\sigma}(t)$.
After that, it can be derived that
\begin{align}\notag
	&\mathbb{E}_{\pi^*}\left[\sum_{a\in \mathcal{A}} \pi^{*}[a|s]A^{\pi}_{\tilde{r}}[s,a]\right]=\mathbb{E}_{\pi^*}\left[\sum_{a\in \mathcal{A}} \pi^{*}[a|s]A^{\pi}_{\tilde{r}}[s,a]\right]+\\ \notag
	&\mathbb{E}_{\pi^*}\left[\sum_{a\in \mathcal{A}}\tilde{\lambda}_{i}(t) \pi^{*}[a|s]A^{\pi}_{\tilde{c}}[s,a]\right]-\mathbb{E}_{\pi^*}\left[\sum_{a\in \mathcal{A}}\tilde{\lambda}_{i}(t) \pi^{*}[a|s]A^{\pi}_{\tilde{c}}[s,a]\right]+\\ \notag
	&\mathbb{E}_{\pi^*}\left[\sum_{a\in \mathcal{A}}\tilde{\mu}_{i}(t) \pi^{*}[a|s]A^{\pi}_{\tilde{p}}[s,a]\right]-\mathbb{E}_{\pi^*}\left[\sum_{a\in \mathcal{A}} \tilde{\mu}_{i}(t)\pi^{*}[a|s]A^{\pi}_{\tilde{p}}[s,a]\right]+\\ \notag
	&\mathbb{E}_{\pi^*}\left[\sum_{a\in \mathcal{A}} \pi^{*}[a|s]\log \pi(t+1)\right]-\mathbb{E}_{\pi^*}\left[\sum_{a\in \mathcal{A}} \pi^{*}[a|s]\log \pi(t+1)\right]\\ \notag
	&=\mathbb{E}_{\pi^*}\left[\sum_{a\in \mathcal{A}} \pi^{*}[a|s]A^{\pi}_{\tilde{r}_{\rm L}}[s,a]\right]-\mathbb{E}_{\pi^*}\left[\sum_{a\in \mathcal{A}}\tilde{\lambda}_{i}(t) \pi^{*}[a|s]A^{\pi}_{\tilde{c}}[s,a]\right]-\\ \notag
	&\mathbb{E}_{\pi^*}\left[\sum_{a\in \mathcal{A}}\tilde{\mu}_{i}(t) \pi^{*}[a|s]A^{\pi}_{\tilde{p}}[s,a]\right]+\mathbb{E}_{\pi^*}\left[\sum_{a\in \mathcal{A}} \pi^{*}[a|s]\log \pi(t+1)\right]-\\ 
	&\mathbb{E}_{\pi^*}\left[\sum_{a\in \mathcal{A}} \pi^{*}[a|s]\log \pi(t+1)\right].
\end{align}
Consider the Taylor expansion of policy $\pi$, is can be obtained that
\begin{align}\notag
	\log\pi_{\theta}(t+1) =& \log\pi_{\theta}(t) +\left.\nabla_{\theta}\log \pi_{\theta}\right|_{\theta=\theta(t)}\left[\theta(t+1)-\theta(t)\right]+\\
	&\gamma\left[\theta(t)\right]\left[\theta(t+1)-\theta(t)\right],
\end{align}
where $\lim_{t\to \infty}\gamma\left[\theta(t)\right]=0$.
Therefore, we can derive that
\begin{align}\notag
	&\mathbb{E}_{\pi^*}\left[\sum_{a\in \mathcal{A}} \pi^{*}[a|s]A^{\pi}_{\tilde{r}}[s,a]\right]=\mathbb{E}_{\pi^*}\left[\sum_{a\in \mathcal{A}} \pi^{*}[a|s]A^{\pi}_{\tilde{r}_{\rm L}}[s,a]\right]-\\ \notag
	&\mathbb{E}_{\pi^*}\left[\sum_{a\in \mathcal{A}}\tilde{\lambda}_{i}(t) \pi^{*}[a|s]A^{\pi}_{\tilde{c}}[s,a]\right]-\mathbb{E}_{\pi^*}\left[\sum_{a\in \mathcal{A}}\mu_{i}(t) \pi^{*}[a|s]A^{\pi}_{\tilde{p}}[s,a]\right]+\\ \notag
	&\mathbb{E}_{\pi^*}\left[\sum_{a\in \mathcal{A}} \pi^{*}[a|s]\log \pi(t+1)\right]-\mathbb{E}_{\pi^*}\left[\sum_{a\in \mathcal{A}} \pi^{*}[a|s]\log \pi(t)\right]-\\ \notag
	&\mathbb{E}_{\pi^*}\left[\sum_{a\in \mathcal{A}} \pi^{*}[a|s]\left.\nabla_{{\theta}}\log \pi_{{\theta}}\right|_{\theta=\theta(t)}\left[\theta(t+1)-\theta(t)\right]\right]-\\ \notag
	&\mathbb{E}_{\pi^*}\left[\sum_{a\in \mathcal{A}} \pi^{*}[a|s]\gamma\left[\theta(t)\right]\left[\theta(t+1)-\theta(t)\right]\right].
\end{align}
Merging partial terms of them can be obtained that
\begin{align}\notag
	&\mathbb{E}_{\pi^*}\left[\sum_{a\in \mathcal{A}} \pi^{*}[a|s]A^{\pi}_{\tilde{r}}[s,a]\right]=\\ \notag
	&\mathbb{E}_{\pi^*}\left\{\sum_{a\in \mathcal{A}} \pi^{*}[a|s]\left[A^{\pi}_{\tilde{r}_{\rm L}}[s,a]-\left.\nabla_{\theta}\log \pi_{\theta}\right|_{\theta=\theta(t)}\left[\theta(t+1)-\theta(t)\right]\right]\right\}+\\ \notag
	&\mathbb{E}_{\pi^*}\left[\sum_{a\in \mathcal{A}} \pi^{*}[a|s]\frac{\log \pi(t+1)}{\log \tilde{\pi}(t)}\right]-\\ \notag
	&\mathbb{E}_{\pi^*}\left[\sum_{a\in \mathcal{A}}\tilde{\lambda}_{i}(t) \pi^{*}[a|s]A^{\pi}_{\tilde{c}}[s,a]\right]-\mathbb{E}_{\pi^*}\left[\sum_{a\in \mathcal{A}} \tilde{\mu}_{i}(t)\pi^{*}[a|s]A^{\pi}_{\tilde{p}}[s,a]\right]-\\ \notag
	&\mathbb{E}_{\pi^*}\left[\sum_{a\in \mathcal{A}} \pi^{*}[a|s]\gamma\left[\theta(t)\right]\left[\theta(t+1)-\theta(t)\right]\right].
\end{align}
Using the transformation and the performance difference theorem, it can be obtained that
\begin{align}
	A^{\pi}_{r_{\rm L}}[s,a]=\left.\nabla_{\theta}\log \pi_{\theta}\right|_{\theta=\theta(t)}\left[\theta(t+1)-\theta(t)\right],
\end{align}
\begin{align}\notag
	&\mathbb{E}_{\pi^*}\left[\sum_{a\in \mathcal{A}}\tilde{\lambda}_{i}(t) \pi^{*}[a|s]A^{\pi}_{\tilde{c}}[s,a]\right]=\\
	&\mathbb{E}_{\pi^{*}}\left[\lambda_{i}(t+1)c[s_{\sigma}(t+1),a_{\sigma}(t+1)]\right]-\mathbb{E}_{\pi^*}\left[\lambda_{i}(t)c[s_{\sigma}(t),a_{\sigma}(t)]\right],\\ \notag
	&\mathbb{E}_{\pi^*}\left[\sum_{a\in \mathcal{A}}\tilde{\mu}_{i}(t) \pi^{*}[a|s]A^{\pi}_{\tilde{p}}[s,a]\right]=\\
	&\mathbb{E}_{\pi^{*}}\left[\mu_{i}(t+1)p[s_{\sigma}(t+1),a_{\sigma}(t+1)]\right]-\mathbb{E}_{\pi^*}\left[\mu_{i}(t)p[s_{\sigma}(t),a_{\sigma}(t)]\right].
\end{align}
It can be then derived that
\begin{align}\notag
	&\mathbb{E}_{\pi^*}\left[\sum_{a\in \mathcal{A}} \pi^{*}[a|s]A^{\pi}_{\tilde{r}}[s,a]\right]=\mathbb{E}_{\pi^*}\left\{\sum_{a\in \mathcal{A}} \pi^{*}[a|s]\left[A^{\pi}_{\tilde{r}_{\rm L}}[s,a]-A^{\pi}_{r_{\rm L}}[s,a]\right]\right\}-\\ \notag
	&\mathbb{E}_{\pi^{*}}\left[\lambda_{i}(t+1)c[s_{\sigma}(t+1),a_{\sigma}(t+1)]\right]+\mathbb{E}_{\pi^*}\left[\lambda_{i}(t)c[s_{\sigma}(t),a_{\sigma}(t)]\right]-\\ \notag
	&\mathbb{E}_{\pi^{*}}\left[\mu_{i}(t+1)p[s_{\sigma}(t+1),a_{\sigma}(t+1)]\right]+\mathbb{E}_{\pi^*}\left[\mu_{i}(t)p[s_{\sigma}(t),a_{\sigma}(t)]\right]+\\ \notag
	&\mathbb{E}_{\pi^*}\left[\sum_{a\in \mathcal{A}} \pi^{*}[a|s]\frac{\log \pi(t+1)}{\log \pi(t)}\right]-\mathbb{E}_{\pi^*}\left[\sum_{a\in \mathcal{A}} \pi^{*}[a|s]\gamma\left[\theta(t)\right]\left[\theta(t+1)-\theta(t)\right]\right].
\end{align}
Therefore, it can be obtained that
\begin{align}\notag
	&\frac{1}{T}\sum_{t=0}^{T-1}\mathbb{E}_{\pi^*}\left[\sum_{a\in \mathcal{A}} \pi^{*}[a|s]A^{\pi}_{\tilde{r}}[s,a]\right]\leq\\ \notag
	&\frac{1}{T}\sum_{t=0}^{T-1}\mathbb{E}_{\pi^*}\left\{\sum_{a\in \mathcal{A}} \pi^{*}[a|s]\left[A^{\pi}_{\tilde{r}_{\rm L}}[s,a]-A^{\pi}_{r_{\rm L}}[s,a]\right]\right\}-\frac{\tilde{\lambda}\bar{c}}{T}-\frac{\tilde{\mu}\bar{p}}{T}+\\ \notag
	&\frac{\log NFC\tau_{\max}}{\eta_1T}-\frac{\gamma\left[\theta(T)\right]\bar{\theta}_{T}}{T},
\end{align}
where the second, third, and fourth term in the right-hand side use the same derivation process in Appendix \ref{apdx:ProofTheorem2}.
The last term has the similar expression as that of the second and third term, so it is derived using the same process.
Using the expansion of $A^{\pi}_{r}[s,a]$ in \eqref{eq:policyparagradient} of Theorem 2 to $A^{\pi}_{\tilde{r}}[s,a]$, it can be derived that
\begin{align}\notag
	&\frac{1}{T}\sum_{t=0}^{T-1}\mathbb{E}_{\pi^*}\left\{\sum_{a\in \mathcal{A}} \pi^{*}[a|s]\left[A^{\pi}_{\tilde{r}_{\rm L}}[s,a]-A^{\pi}_{r_{\rm L}}[s,a]\right]\right\}\leq\\ \notag
	&\frac{\mathbb{E}_{\pi^*}\left[\log\frac{\tilde{\pi}_{T}}{\pi_{T}}\right]}{\eta_1T}+\frac{1}{\eta_1T}\sum_{t=0}^{T-1}\mathbb{E}_{\pi^*}\left\{\sum_{a\in \mathcal{A}} \pi^{*}[a|s]\left[\log \tilde{Z}(s)-\log Z(s)\right]\right\}\\ \notag
	&=\frac{\mathbb{E}_{\pi^*}\left[\log\frac{\tilde{\pi}_{T}}{\pi_{T}}\right]}{\eta_1T}+\frac{1}{T}\left\{\mathbb{E}_{\pi^{*}}\left[\tilde{r}[s_{\sigma}(T),a_{\sigma}(T)]\right]-\mathbb{E}_{\pi^*}\left[r[s_{\sigma}(T),a_{\sigma}(T)]\right]\right\} \\ \notag
	&+  \frac{1}{T}\left\{\mathbb{E}_{\pi^{*}}\left[\tilde{\lambda}_{i}(T)\tilde{c}[s_{\sigma}(T),a_{\sigma}(T)]\right]-\mathbb{E}_{\pi^*}\left[\lambda_{i}(T)c[s_{\sigma}(T),a_{\sigma}(T)]\right]\right\}\\ \notag
	&+ \frac{1}{T}\left\{\mathbb{E}_{\pi^{*}}\left[\tilde{\mu}_{i}(T)\tilde{p}[s_{\sigma}(T),a_{\sigma}(T)]\right]-\mathbb{E}_{\pi^*}\left[\mu_{i}(T)p[s_{\sigma}(T),a_{\sigma}(T)]\right]\right\}
\end{align}
It should be noted that the third and fourth term of the above expressions are affected by the difference of number of subcarriers and power.
Suppose the third term is denoted by $\delta_c$ and the fourth term is denoted by $\delta_p$, 
it can be obtained that
\begin{align}
	&\frac{1}{T}\sum_{t=0}^{T-1}\mathbb{E}_{\pi^*}\left[\sum_{a\in \mathcal{A}} \pi^{*}[a|s]A^{\pi}_{\tilde{r}}[s,a]\right]\leq \frac{g(N,F,C,\tau_{\max})+q(\delta_c,\delta_p)}{T},
\end{align}
where
\begin{align}\notag
	q(\delta_c,\delta_p) = \frac{\mathbb{E}_{\pi^*}\left[\log\frac{\tilde{\pi}_{T}}{\pi_{T}}\right]}{\eta_1}+\tilde{r}_T-2\bar{r}_T+\delta_c+\delta_p-2\bar{\lambda}\bar{c}-2\bar{\mu}\bar{p}-\\
	\gamma\left[\theta(T)\right]\bar{\theta}_{T}+\frac{\mathbb{E}_{\pi}\left(\lambda_i^2(T)\right)-\epsilon_{\lambda} }{2\eta_2}+\frac{\mathbb{E}_{\pi}\left(\mu_i^2(T)\right)-\epsilon_{\mu} }{2\eta_3}
\end{align}
It should be noted that $q(\delta_c,\delta_p)$ is still a finite value.
Therefore, as $T$ approaches infinity, the optimality gap goes to zero.

\end{document}